\documentclass[fleqn,usenatbib]{mnras}

\usepackage{newtxtext,newtxmath}
\usepackage{adjustbox}
\usepackage{orcidlink}
\usepackage[utf8]{inputenc}

\usepackage[T1]{fontenc}

\DeclareRobustCommand{\VAN}[3]{#2}
\let\VANthebibliography\thebibliography
\def\thebibliography{\DeclareRobustCommand{\VAN}[3]{##3}\VANthebibliography}

\usepackage{graphicx}	
\usepackage{amsmath}	

\title[North-PHASE: YSO Variability Survey]{North-PHASE: Studying Periodicity, Hot Spots, Accretion Stability and Early Evolution in young stars in the northern hemisphere}

\author[Sicilia-Aguilar et al.]{A. Sicilia-Aguilar$^{1}$\thanks{E-mail: asiciliaaguilar@dundee.ac.uk}\thanks{Based on observations made with the JAST80 telescope at the Observatorio Astrof\'{i}sico de Javalambre, in Teruel, owned, managed and operated by the Centro de Estudios de F\'{i}sica del Cosmos de Arag\'{o}n.}\orcidlink{0000-0002-8421-0851},
R. S. Kahar$^{1}$,
M.E. Pelayo-Bald\'{a}rrago$^{2,1,3}$,
V. Roccatagliata$^{4,5}$,
D. Froebrich$^{6}$\orcidlink{0000-0003-4734-3345},\newauthor
F.J.  Galindo-Guil$^{7}$\orcidlink{0000-0003-4776-9098},
J. Campbell-White$^{8}$,
J.S. Kim$^{9}$\orcidlink{0000-0001-6072-9344},
I. Mendigut\'{i}a$^{10}$,
L. Schlueter$^1$,
P. S. Teixeira$^1$,\newauthor
S. Matsumura$^{1}$,
M. Fang$^{11}$, 
A. Scholz$^{12}$,
P. \'{A}brah\'{a}m$^{13,14,15}$, 
A. Frasca$^{16}$,
A. Garufi$^5$,
C. Herbert$^{6}$,\newauthor 
\'{A}. K\'{o}sp\'{a}l$^{14,15,17}$, 
and C.F. Manara$^{8}$
\\
$^{1}$SUPA, School of Science and Engineering, University of Dundee, Nethergate, DD1 4HN, Dundee, UK\\
$^{2}$Departamento de F\'{i}ica Te\'{o}rica, Facutad de Ciencias, Universidad Aut\'{o}noma de Madrid, 28049 Cantoblanco, Madrid, Spain\\
$^{3}$Universidad Nacional Mayor de San Marcos. Facultad de Ciencias F\'{i}sicas. Unidad de Postgrado. Ciudad Universitaria, Lima 15801, Peru\\
$^{4}$ Alma Mater Studiorum, Universit\'{a} di Bologna, Dipartimento di Fisica e Astronomia (DIFA), Via Gobetti 93/2, 40129 Bologna, Italy\\
$^{5}$INAF-Osservatorio Astrofisico di Arcetri, Largo E. Fermi 5, 50125 Firenze, Italy\\
$^{6}$Centre for Astrophysics and Planetary Science, School of Physics and Astronomy, University of Kent, Canterbury CT2 7NH, UK\\
$^{7}$Centro de Estudios de F\'isica del Cosmos de Arag\'{o}n (CEFCA), Plaza San Juan 1, 44001 Teruel, Teruel, Spain.\\
$^{8}$European Southern Observatory, Karl-Schwarzschild-Strasse 2, 85748 Garching bei M\"unchen, Germany\\
$^{9}$Steward Observatory, University of Arizona, 933 N. Cherry Ave., Tucson, AZ 85721-0065, USA\\
$^{10}$Centro de Astrobiolog\'{i}a (CAB), CSIC-INTA, Camino Bajo del Castillo s/n, 28692, Villanueva de la Ca\~{n}ada, Madrid, Spain\\
$^{11}$Purple Mountain Observatory, Chinese Academy of Sciences, 10 Yuanhua Road, Nanjing 210023, China\\
$^{12}$SUPA, School of Physics and Astronomy, University of St Andrews, North Haugh, St Andrews KY16 9SS, Scotland, UK\\
$^{13}$Institute for Astronomy (IfA), University of Vienna, T\"urkenschanzstrasse 17, A-1180 Vienna, Austria\\
$^{14}$Konkoly Observatory, HUN-REN Research Centre for Astronomy and Earth Sciences, MTA Centre of Excellence, Konkoly Thege Mikl\'os \'ut 15-17, 1121 Budapest, Hungary\\
$^{15}$Institute of Physics and Astronomy, ELTE E\"otv\"os Lor\'and University, P\'azm\'any P\'eter s\'et\'any 1/A, 1117 Budapest, Hungary\\
$^{16}$INAF -- Osservatorio Astrofisico di Catania, Via S. Sofia 78, 95123 Catania, Italy\\
$^{17}$Max Planck Institute for Astronomy, K\"onigstuhl 17, 69117 Heidelberg, Germany\\
}

\date{Accepted 2024 June 24. Received 2024 June 14; in original form 2024 March 11}

\pubyear{2024}

\begin{document}
\label{firstpage}
\pagerange{\pageref{firstpage}--\pageref{lastpage}}
\maketitle

\begin{abstract}
We present the overview and first results from the North-PHASE Legacy Survey, which follows six young clusters for five years, using the 2 deg$^2$ FoV of the JAST80 telescope from the Javalambre Observatory (Spain). North-PHASE investigates stellar variability on timescales from days to years for thousands of young stars distributed over entire clusters. This allows us to find new YSO, characterise accretion and study inner disk evolution within the cluster context. Each region (Tr~37, Cep~OB3, IC~5070, IC~348, NGC~2264, and NGC~1333) is observed in six filters (SDSS griz, u band, and J0660, which covers H$\alpha$), detecting cluster members as well as field variable stars. Tr~37 is used to prove feasibility and optimise the variability analysis techniques. In Tr~37, variability reveals 50 new YSO, most of them proper motion outliers. North-PHASE independently confirms the youth of astrometric members, efficiently distinguishes accreting and non-accreting stars, reveals the extent of the cluster populations along Tr37/IC~1396 bright rims, and detects variability resulting from rotation, dips, and irregular bursts. The proper motion outliers unveil a more complex star formation history than inferred from Gaia alone, and variability highlights previously hidden proper motion deviations in the surrounding clouds. We also find that non-YSO variables identified by North-PHASE cover a different variability parameter space and include long-period variables, eclipsing binaries, RR Lyr, and $\delta$ Scuti stars. These early results also emphasize the power of variability to complete the picture of star formation where it is missed by astrometry.
\end{abstract}

\begin{keywords}
stars: variables: T Tauri, Herbig Ae/Be -- stars: variables: general --  open clusters and associations: individual: Tr~37, NGC~2264, NGC~1333, Cep~OB3, IC~5070, IC~348 -- stars: formation -- surveys -- techniques: photometric
\end{keywords}



\section{Introduction}

Young stars are born in clusters and acquire most of their mass during the protostellar phase \citep{lada03}. Due to angular momentum conservation, stars emerge surrounded by protoplanetary disks. For a few-Myr, the disks accrete onto the stars and eventually form planetary systems, their properties shaped by accretion. Both the angular momentum evolution of the system and the matter transport through the planet-forming region depend on the accretion process, channeled onto the star via the magnetosphere for most low-mass stars \citep{koenigl91,hartmann_accretion_2016}. The star-disk interaction affects the stellar rotation, and the link between rotation and stellar activity \citep{bouvier_coyotes_1995} also means that accretion can influence the system's habitability. The accretion rate evolution also determines the types of planets formed \citep[e.g.][]{matsumura21} and, since the magnetospheric disk truncation \citep{donati_magnetic_2019} may limit planet migration \citep{Lin96}, the inner disk evolution is crucial to understand the orbital distribution of planets. 
Despite its importance, the main limitation to study the star-disk connection is that its spatial scales range from a few stellar radii \citep[the size of the magnetospheres, e.g.][]{bouvier_magnetospheric_2006,johnstone14}, to fractions of au in the inner planet-forming region, which are beyond imaging reach, especially for solar analogs, most of which are very faint.

Time resolution is a powerful tool to investigate tiny structures. In fact, \textit{"using time to map space"} is the only feasible method to study small-scale processes and structures for statistically significant numbers of sources.
Young T Tauri stars (TTS) are variable by definition \citep{joy45}, so variability is one of the most reliable ways to find young stellar objects \citep[YSO;][]{briceno01}. Variability in YSO is caused by accretion, occultations by circumstellar material (from the disk, accretion columns, or disk winds), and hot and cold spots on the stellars surface \citep{herbst94,bozhinova16,fischer22,herbert23}. The innermost disk structure and star-disk alignment can also be probed by tracking extinction events and their periodicity \citep{roccatagliata11misalignment}. 

The variability timescales reveal the location of the physical processes that produce it \citep[e.g.][]{lightfoot89,eiroa02,hamilton05,sicilia20j1604}. Structures at the stellar surface and up to the disk corotation radius vary due to stellar rotation on timescales of days \citep{bouvier_magnetospheric_2007,cody14,rebull20,froebrich21}, while the inner disk timescales depend on the Keplerian period at each radius, ranging from days to years \citep[e.g.][]{fischer22}. Accreting stars show stronger variations and even outbursts \citep{froebrich23}, but activity, spots and even occultations by circumstellar matter also happen in stars that have ceased to accrete \citep{scholz19}. Thus variability reveals the entire YSO population, unveiling the cluster structure independently of the astrometric properties of its members, and including objects missed by Gaia. 

Most YSO studies so far concentrate on small areas due to field of view (FoV) limitations, but Gaia shows that even well-known clusters extend far beyond their classical limits \citep{roeser19,meingast19,meingast21,mendigutia22,tarricq22} and often have multiple populations \citep{roccatagliata20,teixeira20,pelayo23,szilagyi23}. While the periphery of young clusters is largely unexplored, star formation is found to be more complex than previously thought \citep{wright22}. The cluster outskirts are key to understand star formation \citep{smilgys17,mendigutia22} due to their different initial conditions, and their evolution and interaction with other systems \citep{jerabkova21}. The star-cluster connection is also critical for the evolution of planetary systems. The cluster environment (e.g. sparse vs dense regions, UV irradiation) affects the initial mass function \citep{hsu13}, the sizes of the disks \citep{dejuan12,eisner18}, the disk lifetime \citep{fang13}, and the planetary system properties \citep{kruijssen20}. Finding astrometric outliers among clusters is also important to fully characterise disk evolution \citep{bertini23}. 
Indeed, early interactions also played a role for the young Solar System \citep{kenyon04,mamajek15,delafuente18,bailerjones22}, so exploring a variety of complete young clusters is the best way to put the Solar System in context.

Here, we present the initial results of the North-PHASE project, using the first four months of data and one of the clusters, Tr~37, as a test bench to define and construct the tools needed for variability tracking, and investigating what variability surveys can add to our knowledge, even in very well-known regions. Section \ref{survey-sect} describes the North-PHASE project. The data acquisition, calibration, and variability analysis tools are described in Section \ref{data-sect}. The initial results on the benchmark cluster are presented in Section \ref{tr37-sect}. The conclusions of this first part of the project, including a summary of results, overview of the legacy values, and future plans, are given in Section \ref{conclu}.

\begin{figure*}
    \centering
    \begin{tabular}{cc}
       \includegraphics[trim=1.8cm 0 0.8cm 0, width=9.5cm]{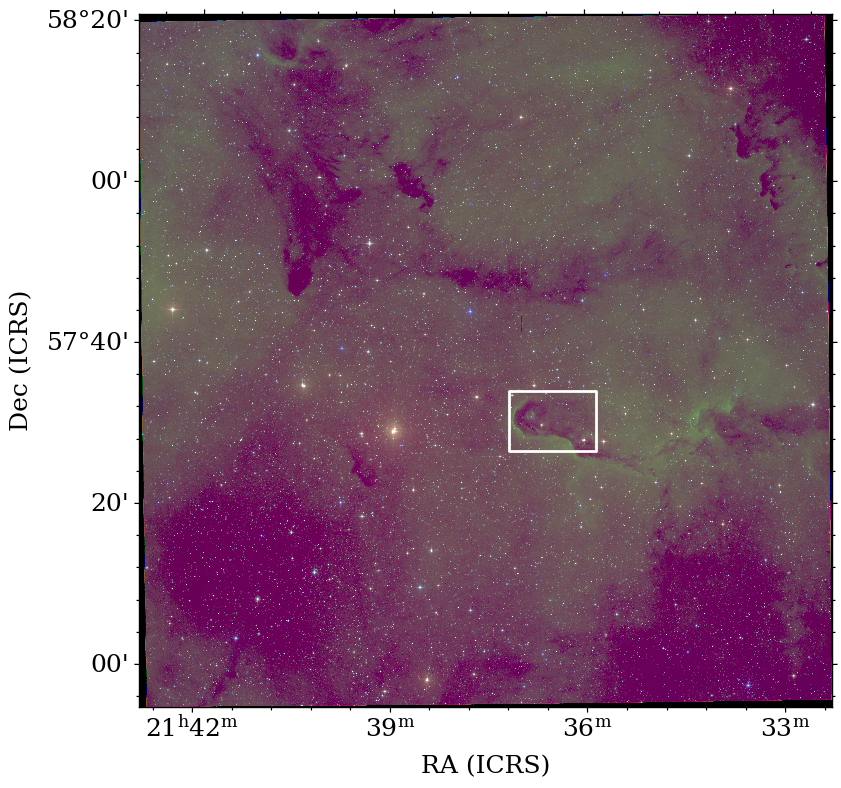} &
    \includegraphics[trim=0 -8cm 0 0.5cm, width=7.5cm]{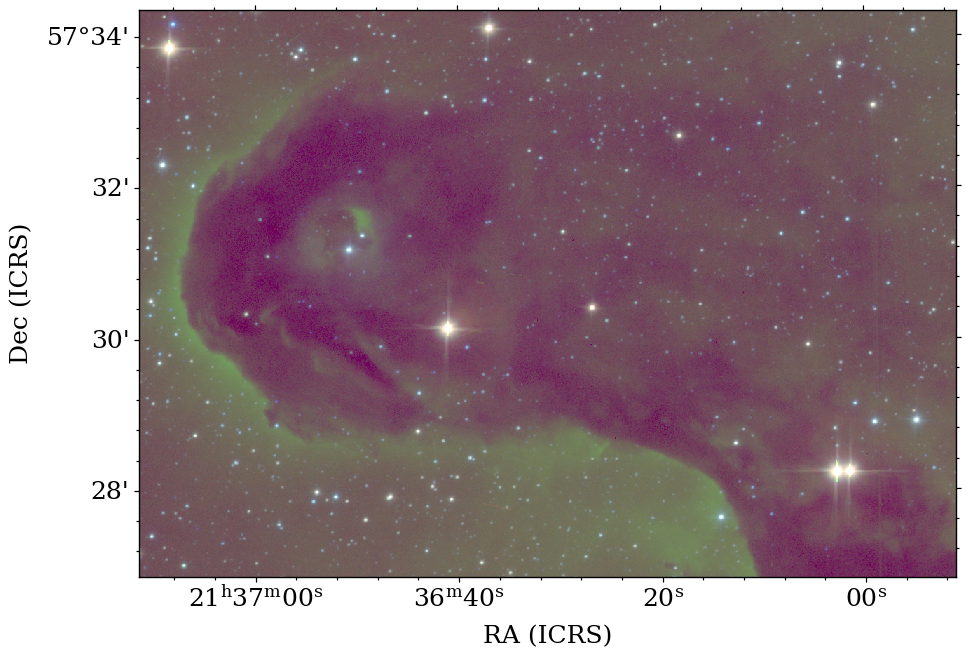}  \\
    \end{tabular}
     \caption{A view of Tr~37 according to North-PHASE in full (left) and zoomed around the white box surrounding the IC1396A globule (right). Red corresponds to zSDSS, green corresponds to J0660 (H$\alpha$) and blue corresponds to gSDSS. All images, shown in log scale, are combination of 9 individual observations.  }
    \label{fig:tr37-view}
\end{figure*}

\section{The North-PHASE survey}
\label{survey-sect}

\begin{table*}
    \centering
\resizebox{\textwidth}{!}{%
\begin{tabular}{lcccccccccccccc}
    \hline
     Cluster & RA & Dec & D & Age & t$_{g}$ & t$_{r}$ & t$_{i}$ & t$_{z}$ & t$_{u}$ & t$_{J0660}$ & Nr. & Nr. Stars &  Dates & Nr. Varia. \\
      & (ICRS) & (ICRS) & (pc) & (Myr) & (s)  & (s) & (s) & (s) & (s) & (s) & Obs. & (rSDSS) & (JD, d) & (4 indices)\\
     \hline \hline
    IC~348 & 03:43:39 & +32:17:53 & 313 & 2-4$^a$  & 7 & 7 & 4 & 7 & 75 & 40 & 51 & 5~211 & 2459979-2460025 & 102 \\
    NGC~1333 & 03:30:50  & +31:19:41 & 292 & 1-2$^a$  & 7 & 7 & 4 & 7 & 75 & 40   & 54 & 4~983 & 2459979-2460025   & 59\\      
    NGC~2264 & 06:40:51 &  +09:23:20 & 730 & 1-3$^a$ & 30  & 30 & 15 & 30 & 300 & 150 & 73 & 24~567 & 2459981-2460046 & 593 \\
    IC~5070 & 20:52:05  & +44:06:39 & 800 & 1-3$^b$  & 30  & 30 & 15 & 30 & 300 & 150  & 39 & 30~376 & 2460039-67 & 317\\
    Tr~37 & 21:37:26  & +57:38:50 & 925$^c$ & 1-4$^c$ & 30  & 30 & 15 & 30 & 300 & 150  & 31 & 89~680 & 2460040-67 & 894 \\
    Cep~OB3 & 22:58:11 & +62:41:06 & 825 & 2$^d$  & 30  & 30 & 15 & 30 & 300 & 150  & 9 & 32~625 & 2459979-81 & --  \\
    \hline
    \end{tabular}
    }
    \caption{Summary of the observations and current results for North-PHASE. Central coordinates, distances (estimated from Gaia) and ages (see references below) are given, plus filters and exposure times (in s) for each cluster as well as a summary including number of nights analysed, number of stars detected at least 6 times in the rSDSS filter, and variable stars of any type identified (following the methodology of Section \ref{timeres-sect}, which requires variability in at least 4 indices and g<19 mag). The uJAVA filter is only done on dark nights (about 1 out of every 5). Cep~OB3 does not contain enough data to look into variability yet. The references for the cluster ages are $^a$\citet{richert18}, $^b$\citet{froebrich21}, $^c$\citet{pelayo23}, and $^d$\citet{getman12}.}
    \label{tab:obstable}
\end{table*}

The "Periodicity, Hot Spots, Accretion Stability and Early Evolution in young stars in the northern hemisphere" survey (North-PHASE) is a legacy survey aimed at \textit{"using time to map space"} in young stars via variability. The project targets six clusters with stars aged 1-5 Myr (Tr~37, NGC~2264, NGC~1333, IC~348, IC~5070, and Cep~OB3), containing altogether several thousands of YSO, using the Javalambre Observatory 83~cm JAST80 telescope in Spain\footnote{https://oajweb.cefca.es/telescopes/jast80}.  The clusters were selected to ensure a range of ages (from 1-5 Myr), a diversity of star forming histories (low-mass vs high-mass star forming regions) and for being visible from Javalambre for at least 7 months per year. North-PHASE started in February 2023, and will continue until January 2028, producing five years of time-resolved data that will be updated periodically. Our aim is to perform a large-scale, large-FoV, systematic survey to find YSO, explore their variability, investigate accretion over time and its dependency on stellar mass, obtain statistics on accretion outbursts and their role in star and planet formation, and study the evolution of stars and disks within the cluster context.

The first advantage of North-PHASE is the 2~deg$^2$ FoV of the T80Cam camera\footnote{\href{https://oajweb.cefca.es/telescopes/t80cam}{https://oajweb.cefca.es/telescopes/t80cam}}, which allows us to cover each cluster in one shot. The large FoV and variable cadence (being sensitive to variability from days to years) are critical to identify YSO in an unbiased way, including the clusters' outskirts beyond the "classical" cluster boundaries. This provides an advantage in extension, depth, and spatial resolution compared to similar surveys with smaller FoV \citep[such as HOYS,][]{froebrich18,froebrich24}. Figure \ref{fig:tr37-view} shows the large FoV, plus a zoom in a high-extinction region to reveal the depth of the images. The pixel scale is 0.55"/pixel.
In addition, by using five of the broad band filters within the J-PLUS filter system\footnote{\href{https://www.j-plus.es/survey/instrumentation}{https://www.j-plus.es/survey/instrumentation}} (g,r,i, z Sloan SDSS filters), the u-band filter (uJAVA), and the J0660 narrow-band filter (which covers from $\sim$6510-6690\AA\,, including the H$\alpha$ line), we can reliably distinguish the processes happening in stars and disks.  The six filters also give us an advantage compared to other variability surveys such as the Zwicky Transient Facility \citep[ZTF]{bellm19}.
The observations are planned with variable cadence, with the main cadence being one to three days during periods from 10 days to a month, and observations blocks spaced by a couple of months, depending on visibility, to target the key variability timescales in young stars \citep[e.g.][]{fischer22}.
Details of the pointings are given in Table \ref{tab:obstable}. 

North-PHASE aims to achieve a complete census of YSO aged 1-5 Myr, down to 0.15-0.3 M$_\odot$, classifying variability types and mechanisms as well as measuring stellar parameters.
The 5 years of observations will allow us to measure how accretion and rotation are linked through this age interval for stars with different masses. Changes in accretion on these timescales provide information on the stability of the star-disk connection, and the role of perturbations by planets or companions in the mass transport through the inner disk, and the variability of the magnetic field.
Thanks to the u-band and H$\alpha$ filters, we will trace accretion rates and how they vary. Combining the information on stellar rotation, spot-like variability, and extinction events, we will determine magnetosphere sizes and explore the connection between the star and the inner planet-forming regions and explore longer-term changes in accretion related to changes in the inner disk and magnetic cycles in young stars.

In addition, North-PHASE will also provide very deep combined images and detect other types of variable field stars, so its legacy value expands beyond the star formation community to general stellar variability, stellar properties and the requirements to use the JAST80 in galactic observations. The combination of long and short cadences renders it well-suited for pinpointing fast-moving sources, such as
objects within our solar system \citep{cortes19,racero22}.
Finally, the large FoV and diversity of pointings as well as the number of visits per year (69) will also allow us to quantify the effects of satellite constellations trails on datasets such as ours, which has been done so far mostly using simulations \citep{hainaut20,bassa22}.

\section{Observations and data calibration}
\label{data-sect} 

\subsection{Observations}

For this first analysis of the North-PHASE programme, we studied the data acquired between the start of the programme in February 2023, and May 2023. The observations consist of 3 exposures per night for filters g, r, i, z, and J0660 (H$\alpha$), acquired with variable cadence. On dark nights (about 20\% of the time), a further set of 3 observations is recorded using the near-UV uJAVA filter. Due to a telescope glitch, some datasets had 4 observations. Exposure times depend on the filter, and are halved for the nearby regions IC~348 and NGC~1333 (see Table \ref{tab:obstable}). All clusters are observed following a similar pattern but, due to their coordinates and visibility, at the time of this analysis there is still a very substantial difference between the number of datapoints available for each region, with NGC~2264 so far having over 70 datasets in filters grizH$\alpha$, while this number is reduced to 9 for Cep~OB3.

The 3 nightly repetitions are meant to achieve a deeper photometric coverage by combining the data. Used individually, the nearly-simultaneous images allow us to test the data and relative calibration self-consistency, and may also reveal in the future very short timescale variability in certain targets. In the current work, we have explored the result of using the datapoints independently vs averaging them per day. Further details are discussed in Section \ref{timeres-sect}.

\subsection{Relative photometry calibration}

The raw data products delivered to North-PHASE are similar to those provided for the J-PLUS survey\footnote{\href{https://archive.cefca.es/doc/manuals/catalogues_portal_users_manual.pdf}{https://archive.cefca.es/doc/manuals/catalogues\_portal\_users\_manual.pdf}}, including photometric catalogues and basic processed images. 
To produce these catalogues, the data are first reduced using the J-PLUS pipeline\footnote{\href{https://www.j-plus.es/datareleases/data_release_dr3}{https://www.j-plus.es/datareleases/data\_release\_dr3}} \citep{cenarro19,bonoli21}. This includes the basic calibration of the images, including merging the 8 CCDs of the T80Cam detector and performing standard bias, flat fielding, cosmic ray removal  \citep{vandokkum01}, distortion correction, and coordinate calibration. The pipeline then creates the processed catalogues with various types of photometry. Since the pipeline was designed for the detection of galaxies, {\sc sextractor} \citep{bertin96,bertin02,bertin06,bertin11} is first used to derive photometry, but additional aperture photometry is also given for apertures in the range 1-20". 
The pipeline also provides an absolute calibration based on Pan-STARRS objects in the field and performed against the 6" aperture. Although useful for a first-look, the pipeline calibration is optimised for extragalactic fields with very low extinction, and it was found to have offsets as large as 0.2 mag from night to night in our clusters.

For variability studies, a robust relative correction that allows us to compare the magnitude variations from night to night is more important than an absolute flux calibration, as it already allows us to identify and classify the variable stars. Therefore, the main aim at this stage was to optimise the relative photometry calibration. 
We first selected the best aperture, 3" (nominally 2.8"), which, after inspection of the images and comparison of results, was found to offer the best compromise between light collection and avoiding overlap in crowded regions or near nebulosity. We then built customised {\sc Python} routines for the relative calibration and variability index calculation. Cross-matching between catalogues was done using the Starlink Tables Infrastructure Library Tool Set ({\sc STILTS}), which improves code efficiency. {\sc STILTS} was called within the same {\sc Python} program used to calibrate. In general, we use a radius of 0.5" and the {\sc tmatchn} subroutine for cross-matching of objects. The same procedure was used to match the North-PHASE catalogues to the Gaia DR3 survey in the analysis of Tr~37 (see Section \ref{tr37-sect}).
The present calibrations were done against the first observation in each set, which were good nights in terms of seeing and transparency. At this stage, this allows us to get the most of the still-limited datasets available, with the plan set to derive deeper calibrations based on many combined images \citep[as done in][]{froebrich18,evitts20} as the survey advances.

The relative calibration for individual nights follows from a sigma-clipping procedure \citep[similar to][]{sicilia04,sicilia08gmcep,sicilia13}. For each filter and date, we extracted the 3" aperture photometry data, and performed a cross-match with the first observation
The data were clipped to remove dubious points and outliers, requiring a minimum of 7 observations to be considered as a candidate for calibration. 
Further clipping was used to remove variable objects as well as those too faint or too close to the saturation limit, comparing with the magnitudes in the reference night. 

We explored whether there could be differences in the calibration for fainter or brighter stars, but found none. Colour terms are negligible, as expected, since each filter is compared with itself. Therefore, the relative calibration was then derived as the average and standard deviation of the magnitude difference with respect to the reference line, for the clipped stars. The final relative magnitudes were derived by adding the corresponding nightly calibration offset to the 3" photometry. As a final verification, light curves of non-variable objects were examined, revealing no significant correlations. This was further checked in the analysis of Tr~37 (Section \ref{tr37-sect}).

The final relatively-calibrated catalogues were cleaned, removing spurious objects from cloud rims and bleeding patterns by requesting sources to be observed in at least 5 independent images. This may miss extreme variables or faint outbursting stars but, as any other very faint object, we will not be able to study those in detail until more data have been gathered. Future deep combined images will also help to obtain a detailed catalogue of verified detections for these very faint targets. The resulting calibrated tables and variability indices are available via the Javalambre database\footnote{\href{https://archive.cefca.es/catalogues/}{https://archive.cefca.es/catalogues/}}.

As for the dynamical range of the survey, for the distant clusters we find that stars brighter than Gaia g=12.5 mag tend to suffer from saturation, while those fainter than Gaia g=19 mag are often much affected by noise in filters rSDSS, iSDSS, zSDSS, and J0660 (or after $>$17.5 mag, for gSDSS). Objects beyond both ends are excluded from the following analysis. The uJAVA filter is compromised at an even brighter stage, but we will use J0660 to study accretion variability, with uJAVA required to calibrate accretion at less-regular intervals. We also note that these limits refer to individual observations with the exposure times reflected in Table \ref{tab:obstable}. Very faint objects are still detectable by combining the images, even though this reduces the number of time samples. As for addressing saturation, among intermediate-mass stars, we have modified the sequence from the second year of the survey on by including one short exposure out of each set, which strongly reduces the extent of the problem.

\subsection{A flux calibration for Tr~37}
\label{sect:fluxcal}

\begin{table}
    \centering
    \begin{tabular}{llc}
    \hline
    Filter & Correction (mag) & Nr. Sources \\
    \hline\hline
     gSDSS & 1.633$\pm$0.014  & 5546  \\
     rSDSS & 1.816$\pm$0.015  & 9299 \\
     iSDSS & 0.862$\pm$0.021  & 10377  \\
     zSDSS & 1.147($\pm$0.001) + 0.1052($\pm$0.0015)$\times$(r-i) & 12320 \\
     J0660 & 1.411$\pm$0.039  & 9466 \\
     \hline
    \end{tabular}
    \caption{Calibration results for Tr37, relative to the master night JD 2460039.7 (Sect. \ref{sect:fluxcal}). The calibration is given as a zero-point for sources without colour terms, while zSDSS has significant colour terms and thus a more complex transformation. The number of sources used in the fit is also given. 
     \label{tab:tr37-cal}
    }
\end{table}

Although the quality of the absolute calibration does not impair the identification of variable stars, for the purpose of validation and to establish the methodology to follow for the rest of the survey, we also performed an absolute flux calibration for Tr~37. 
An absolute calibration is essential for placing the sources in the HR diagram to measure stellar properties, as well as for deriving absolute accretion rates. The derivation of absolute stellar properties requires good understanding of the variability due to different phenomena (accretion, spots, extinction by circumstellar and interstellar material).

Since the relative calibration is based on the observations taken on JD~2460039.67~d, we also performed the absolute calibration on that dataset. 
To calibrate the broad-band filters, we used data from Pan-STARRS1 \citep{panstarrs16,magnier20,flewelling20}, extracted with {\sc python.astroquery}\footnote{\href{https://ps1images.stsci.edu/ps1_dr2_api.html}{https://ps1images.stsci.edu/ps1\_dr2\_api.html}}. We found the filters to be very similar, and colour terms are only significant for zSDSS, for which a colour term on (r-i), chosen because of the lower level of noise for the red population, substantially improves the fit.
For the narrow-band H$\alpha$ filter J0660, we used data from the INT/WFC Photometric H-Alpha Survey of the Northern Galactic Plane (IPHAS) survey \footnote{\href{https://www.iphas.org/dr2/}{https://www.iphas.org/dr2/}} \citep{barentsen14}. IPHAS also includes data on rSDSS and iSDSS, which we compared with Pan-STARRS1. In particular, for rSDSS, which has negligible colour terms, the correction agrees down to few \%. The correction for iSDSS includes non-negligible colour terms, so it is not directly comparable. As a rule, we take the corrections derived from Pan-STARRS1 because it is also being used for the general Javalambre calibration, it is less noisy, the colour terms are lower, and they include all the 4 main broad-band filters. 

The calibration of the uJAVA filter is critical to determine the accretion rate \citep{gullbring98}. For Tr~37, we used the Johnsons U-band data from \citet{sicilia10}, which had been acquired with the Calar Alto 3.5m telescope using the Large Area Imager for Calar Alto (LAICA) camera\footnote{\href{https://www.caha.es/CAHA/Instruments/LAICA/manual.html}{https://www.caha.es/CAHA/Instruments/LAICA/manual.html}}. Since the uJAVA from the North-PHASE survey and the Johnsons U filter used by LAICA are significantly different, the colour terms are large and we found that the currently available datasets did not have enough depth to guarantee sufficient sources for calibration. We therefore leave the calibration of uJAVA aside for now, to be 
improved with further knowledge of the photometry of individual stars as well as absolute calibrations obtained with the JAST80 telescope.  

The results of the calibration are presented in Table \ref{tab:tr37-cal}.
Saturation for Tr~37 members occurs for r$<$13.3 mag, and the noise level tends to make it very hard to determine variability for objects fainter than r=19.3 mag. For Tr~37 members with extinction A$_V \sim$1.4 mag \citep[see][from now on PB23]{pelayo23}, this ranges from massive T Tauri stars (spectral types F-G) down to M3-M4 types, or a mass of about 0.3 M$_\odot$, although gSDSS is more limited ($\sim$ 0.4 M$_\odot$). This is slightly better than the planned final detection limit of 0.3 M$_\odot$ at 900pc and A$_V\sim$1.5 mag. While the sensitivity in the lower-mass end will increase with combined observations in the future, the fact that most A-type stars are already at least partly saturated is an issue, and it will be mitigated in the future by halving the exposure time of one out of each of the 3 repetitions.

\begin{figure*}
    \centering
   \begin{tabular}{c}
  {\Large {\bf Tr~37}}\\
    \includegraphics[width=16cm]{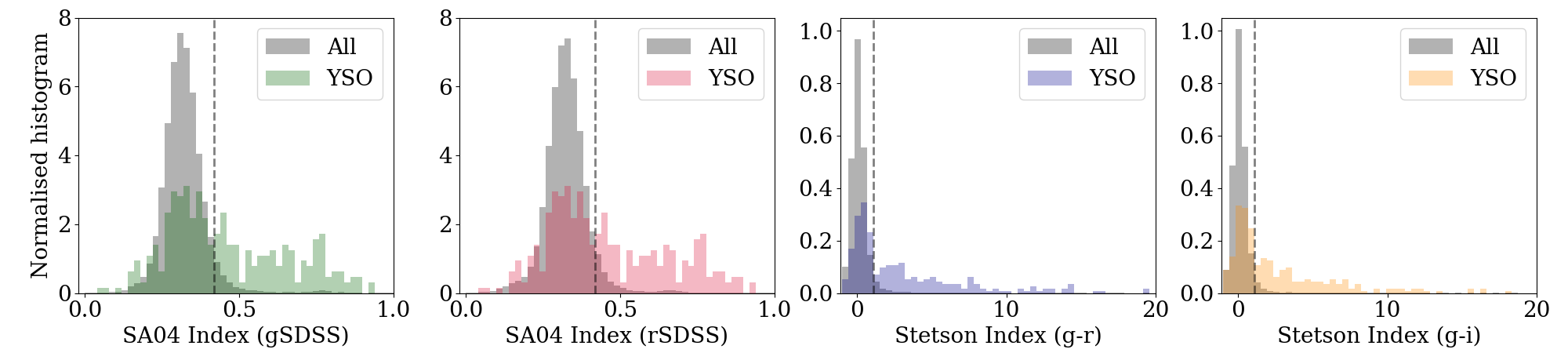}\\
    {\Large {\bf NGC~2264}}\\
   \includegraphics[width=16cm]{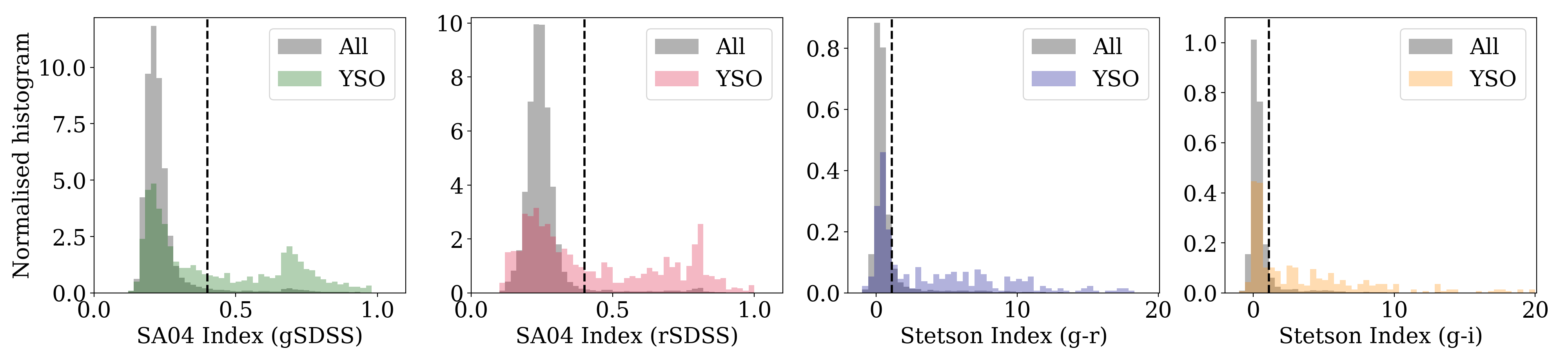}\\
   \end{tabular}
    \caption{Normalised histograms showing how the SA04 and Stetson indices behave differently for known members vs the general field stars. Top: results for Tr~37. Bottom: results for NGC~2264. The YSO were taken from PB23 for Tr~37 (see also Section \ref{tr37-sect}), while for NGC~2264 we use those listed as cluster members by SIMBAD. In both cases, they are compared to the bulk of the population detected in each filter, including members and field stars. The SA04 indices tend to better characterise variables in the clusters even if the observations are partial, although the Stetson indices have the advantage that they are less magnitude-dependent. We observe differences in variability between clusters but, in general, the results for other clusters/other filters are very similar. }
    \label{fig:sa04-analysis}
\end{figure*}

\subsection{Variability analysis and variability indices}
\label{timeres-sect}

Variability indices offer a quick way to detect different types of variable stars based solely on relatively-calibrated data. The core of the North-PHASE programme is detecting stellar variability and using it to classify stars. The first step in this process is to create the light curves for each source, which will be later analysed via variability indices. The data for each source were combined using {\sc STILTS}, incorporated into the {\sc Python} programs used to calculate the variability indices. The data were further cleaned, removing objects that have not been detected a minimum number of times (set to 5) to remove contaminants such as cloud structures as well as mistaken detections that appear on the bleed tracks of saturated objects. All objects were also assigned a set of 'working coordinates' which, for each band, corresponded to the first dataset where the star was detected. The datasets are then considered on a band-by-band basis, or matched to another band.

We explored several variability indices and filter combinations, using the known members in Tr~37 to test them and to set the limits of what should be considered as variable while minimising inclusion of non-variables (more details are given in Section \ref{tr37-sect}). The main indices we considered were the Stetson indices \citep[defined for pairs of filters,][]{stetson96}
and a modification of the simple variability indices from \citet{sicilia04}. 

\begin{figure*}
    \centering
    \begin{tabular}{c}
    {\Large {\bf Tr~37}}\\
    \includegraphics[width=16cm]{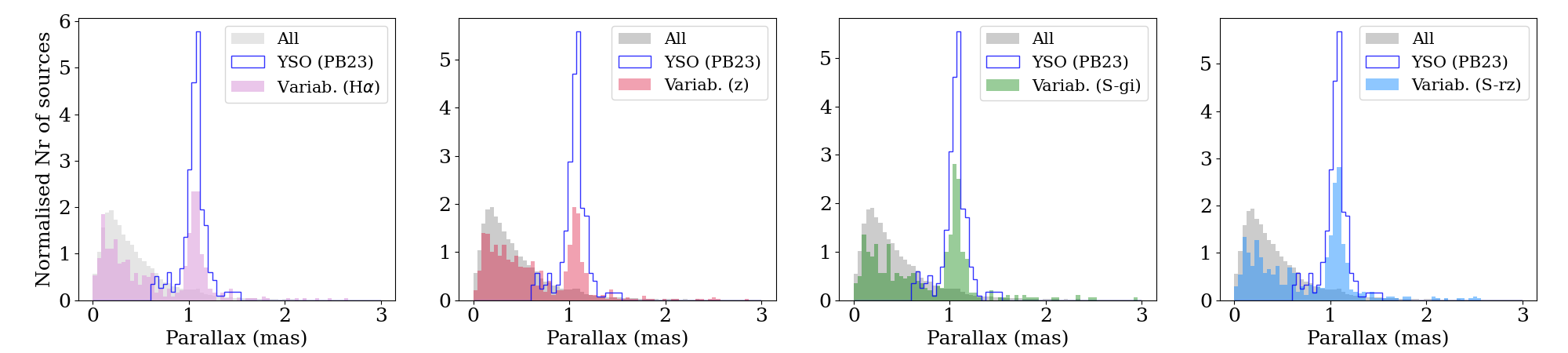}\\
   {\Large {\bf NGC~2264}}\\
    \includegraphics[width=16cm]{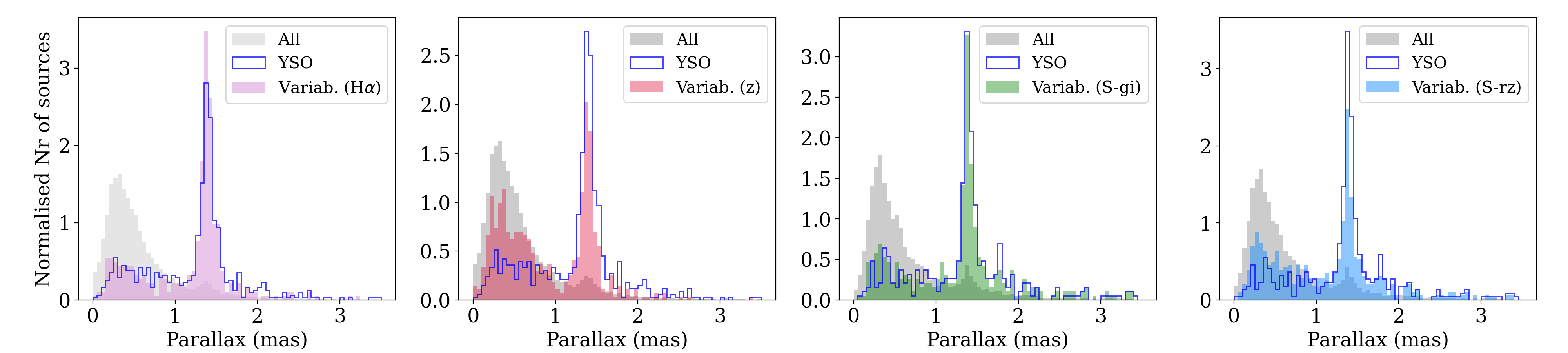}\\
    \end{tabular}
    \caption{Normalised histograms showing the parallax distribution for stars classified as variable, using a single index, compared to the bulk of field stars, and to known YSO. We set  the limits for variability at $>$0.42 for SA04 indices, and $>$1.1 for Stetson indices. The top panels show the results for Tr~37, using the YSO compiled from the literature by PB23. The bottom panels show the results for NGC~2264, with the YSO from SIMBAD. Note that, as shown, both YSO samples contain a non-negligible contamination but even so, for both clusters, the peaks at the cluster parallaxes of $\sim$1.1 and 1.5 mas are evident. }
    \label{fig:parallax-variables}
\end{figure*}

There are various ways to define the Stetson index \citep[see][]{stetson96}. The efficiency of the index to distinguish variables is similar, although the thresholds obviously differ. For each pair of filters $a$ and $b$, the Stetson index $S_{a-b}$ can be calculated as
\begin{equation}
    S_{a-b}=\frac{1}{\sqrt{N(N-1)}} ~~~~ \sum_{i=0}^N sign(P(a-b)_i) \sqrt{|P(a-b)_i|}
 \end{equation}
where
\begin{equation}
  P(a-b)_i = \frac{(m_{a,i}- \overline{m}_a)}{\sigma_{a,i}}   \frac{(m_{b,i}- \overline{m}_b)}{\sigma_{b,i}}.  
\end{equation}
Here, $m_{x,i}$ represents the i-th magnitude in filter $x$, $\overline{m}_x$ is the average magnitude for the same filter, $\sigma_{x,i}$ is the uncertainty in the i-th measurement, and $N$ is the number of observation pairs. There is no rule as to how the filters can be paired, so we tried all combinations of the broad-band filters except uJAVA. We explored pairing individual observations, that is, joining each one of the (typically 3) repetitions of each filter with the corresponding repetitions of the other filter separately, for each one of the nights, as well as using the average of the nightly repetitions. In this case, the uncertainty is taken to be the standard deviation of each set of measurements. We found that the second method offered a better estimate of the uncertainties, and was thus adopted as the standard for the survey's Stetson indices.

The simple variability indices defined by \citet{sicilia04} were based on counting the number of data points that exceeded the uncertainty threshold for a given magnitude as 'variable', and using the percentage of 'variable' points vs the total to assess the variability of a source. For North-PHASE, we introduced a modification, estimating the Gaussian probability of each measurement, assuming that the measured values are normally distributed around the mean with $\sigma$ equivalent to the typical uncertainty for each magnitude range. This variant of the index has the advantage that it will give more weight to objects that have very significant deviations on a small number of nights, which were missed by simple counting and could affect rapid, irregular variables \citep[e.g.,][]{sicilia08gmcep}. The improved SA04 indices also allow to better detect sources that have sustained, lower-level variability, such as those with small and/or cold stellar spots and rotational modulation. To estimate the magnitude-dependent typical uncertainty for each night ($\sigma$), we followed \citet{sicilia04}, binning the nightly dataset in magnitude intervals, and estimating the typical spread compared to the average magnitude. A 3-degree polynomial function was then fitted through the bins to derive the 'uncertainty curve' as a function of magnitude. Each individual magnitude measurement was then compared to the uncertainty curve for its night/dataset, so that the total deviation for the magnitude in filter $a$ for the i-th night ($m_{a,i}$) is
\begin{equation}
    \Delta_{a,i}=\frac{|m_{a,i} - \overline{m}_a|}{\sqrt{2} \times \sigma_{fit,a,i}(m_{a,i})},
\end{equation}
where $\overline{m}_a$ is the average magnitude for that star, and the $\sigma_{fit,a,i}$ function gives the expected (fitted) uncertainty for band $a$ on night $i$, applied to the magnitude $m_{a,i}$. The $\sqrt{2}$ factor is simply the normalisation, so that the deviation for each night will be given by the error function applied to $\Delta{a,i}$, and the final variability index is the normalised sum of all the deviations, i.e.,
\begin{equation}
    SA04_{a}=  \frac{1}{(N-1)} \sum_{i=0}^N erf(\Delta_{a,i}), 
\end{equation}
where N is the total number of observations for that star in $a$ band (N$>$5 due to our requirements). The calculation was done for all broad bands (griz) plus H$\alpha$ (J0660).
We note that the SA04 indices have a slight magnitude dependency, which is normally caused by the polynomial fit overestimating $\sigma$ at the brightest end. If any, this tends to make bright variable stars slightly harder to identify, an issue that we will explore in more detail in the future, with more observations.

As a first test, Fig. \ref{fig:sa04-analysis} shows the difference in index distribution for young stars compared to all detected objects in the field for Tr~37 and NGC~2264, even when a single index is used. For Tr~37, we selected the YSO from PB23, while for NGC~2264, we selected those that are listed as cluster members by the SIMBAD database. Although both clusters have different ages and different variability levels as well as a different number of YSO, the distributions of indices for YSO vs. the field population are very different. We also observe that YSO tend to show a bimodal distribution in the SA04 indices, with objects peaking at $\sim$0.3 (no variability) and $\sim$0.7. As we will see in Section \ref{variability-sect}, this peak is a key in the variability behaviour of YSO and can be explored to distinguish them from other, non-young variables.
Conversely, we find that even if only one index is considered, the distribution of parallaxes for the variable sources is clearly peaked at the clusters distances (see Fig. \ref{fig:parallax-variables}), a further confirmation that the variability indices are suitable for identifying young stars.

Table \ref{tab:obstable} shows the initial results regarding numbers of nights available for each cluster within this preliminary dataset, as well as the number of stars detected (requiring observations in at least 5 nights to call it a detection), and the number of variable stars.

\section{A case study: Tr~37}
\label{tr37-sect}

The Tr~37 cluster \citep{loren75,platais98,contreras02} is one of the best-studied young-star forming regions. It has been targeted by stellar variability studies \citep[with a far smaller scope than North-PHASE, e.g.][]{sicilia04,morales09,meng19}, near-infrared 2MASS and Spitzer surveys \citep{reach04,froebrich05,sicilia06,morales09,Rebull2013}, X-ray \citep{getman07,mercer09,getman12}, spectroscopy \citep{sicilia05cepob2,sicilia06hecto,nakano12,sicilia13,pelayo23}, far-IR Herschel data \citep{sicilia-aguilar_accretion_2015}, and Gaia (PB23). H$\alpha$ photometry \citep{barentsen11} and near-UV data \citep{sicilia10} have been used to measure accretion rates. 

This large collection of data have been used to constrain the properties of hundreds of YSO (e.g. the total number of members and likely members from the literature is 1536; PB23), making it an ideal place to optimize our variable-detecting algorithms.
Of the 1872 members and member candidates listed by PB23, including their new astrometrically-identified ones, 1489 are within the North-PHASE field\footnote{ Some YSO in the extended IC~1396 clouds \citep{barentsen11,nakano12,Rebull2013} fall out of the North-PHASE FoV. }.
Of those, 626 are detected in all filters within the established good-quality limits, and 315 among them have good Gaia DR3 data \citep[defined as those with a Re-normalized Unit Weight Error, RUWE, below 1.4 for a reliable astrometric solution\footnote{In the subsequent sections, 'good astrometric data' corresponds to this RUWE$<$1.4 criterion.};][]{Roccatagliata2020,lindegren21}. This last point is the most restrictive one and, since a large majority of the objects are faint, is satisfied by only 16,878 objects out of 37,882 sources with at least 5 observations per filter (excluding uJAVA). 
The following sections explore the best ways to identify variable stars and discuss the properties of the new variable YSO, exploring for this purpose all objects with reliable Gaia astrometry and ancillary information regarding the presence of protoplanetary disks, accretion, and activity. For the time range considered, the observations of Tr~37 span 47 days, and we are excluding the uJAVA filter (with too few datasets) from the analysis.

\subsection{Examining the accuracy of the variable identification}
\label{basicvaria-sect}

\begin{figure*}
       \centering
       {\Large\bf{Stetson Indices}}\\
    \adjincludegraphics[width=14cm]{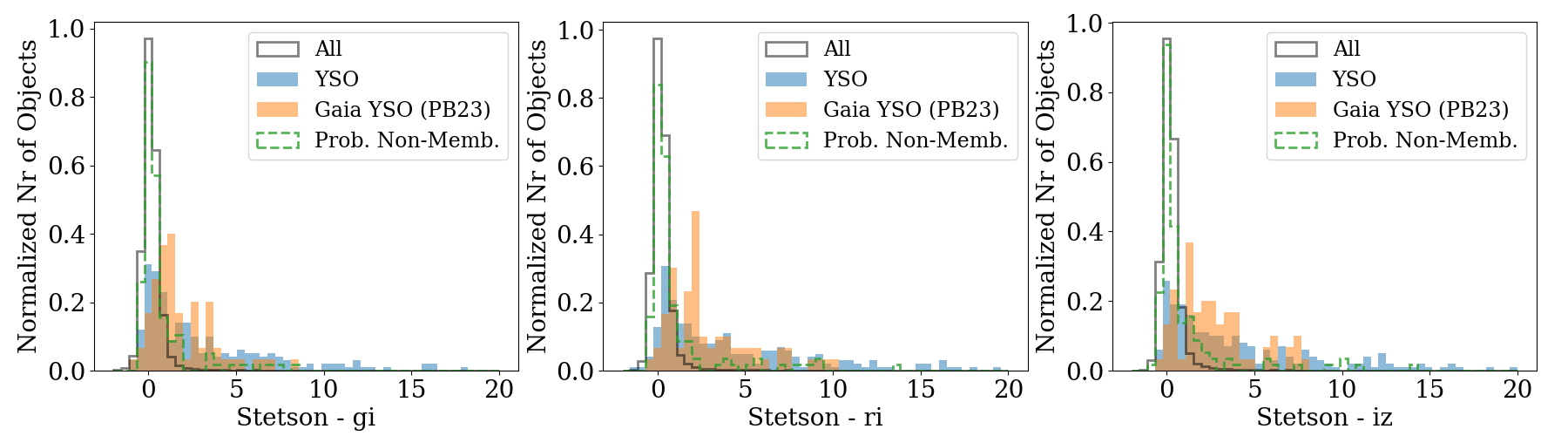}\\
    {\Large\bf{SA04 Indices}}\\
    \includegraphics[width=14cm]{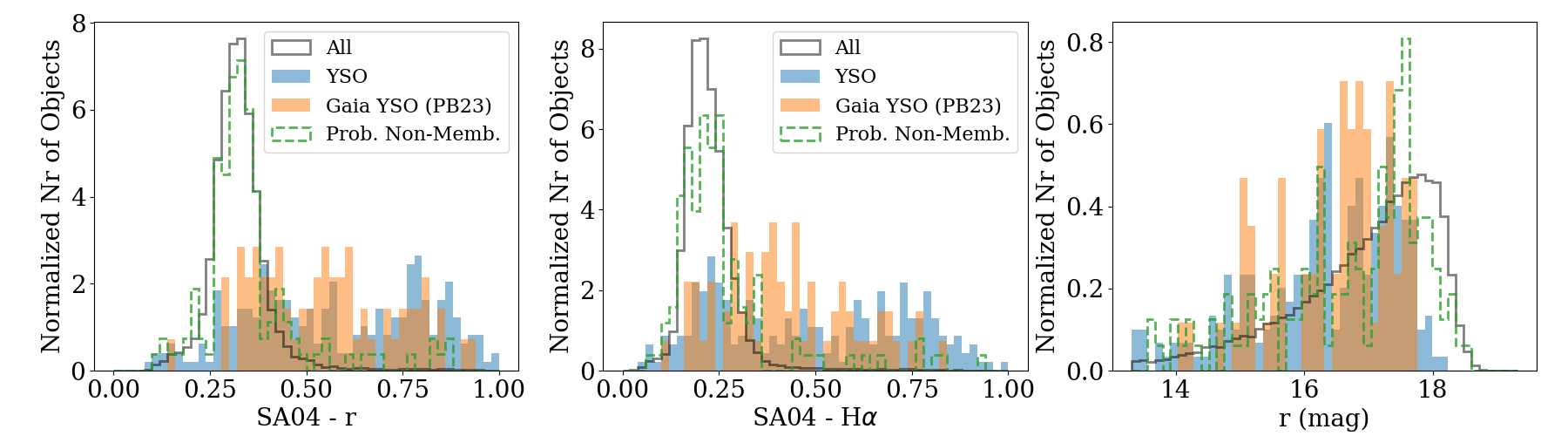}\\
    \caption{Normalised histograms for the variability indices for Tr~37, including the field  population, the highly reliable members from the literature (YSO),  the objects selected by PB23 as astrometric members, and the probable non-members. The differences between YSO and field stars were used to define the variability criteria. While the objects classified as probable non-members are very similar to the field stars, astrometric members are similar to those found by other means.
     The last panel of the SA04 indices block shows the rSDSS magnitude range.}
    \label{fig:index-vs-membership}
\end{figure*}

We first tested the capabilities of North-PHASE for detecting young and variable stars, which requires a uniform sample of members with negligible contamination. To make sure that our analysis was not misled by objects of dubious membership, we first filtered a group of high-confidence members from the original list from PB23. In addition to requesting RUWE$<$1.4, parallax limits between 0.6-1.6 mas were also enforced. This resulted in 321 objects, of which 71 had been identified exclusively by astrometry, among which 315 have data in all North-PHASE bands. 
This group composes our low-contamination sample to test the variability criteria. 
We note that most bona-fide members (confirmed by spectroscopy and IR excesses) do not fall in this category due to the stringent limits, but the main aim with this part is to test the methodology. Studying the variability of the known members as a whole will come as a second objective, once the criteria are well-established.

Similarly, we extracted a collection of probable non-members from the initial potential member list, composed of those with a good astrometric solution but parallax outwith the member range. This is used as a control group for variability purposes, since they are in general stars that share many characteristics with the high-probability members in terms of magnitudes and colours (and hence were at some point suggested to be cluster members), but are most likely unrelated background or foreground objects. There are a total of 134 sources in this class.

We first checked the correlation between indices for YSO and sources that have been rejected as cluster members. The results are shown in Appendix \ref{app-figs}, Fig. \ref{fig:compare-index-memb-nonmemb}. Besides showing a clear difference in variability indices between members and non-members, it also reveals a strong correlation between different Stetson and SA04 indices, as expected if they accurately trace variability. The correlation is not linear, which is good because it means each index is more sensitive to different types of variability.

We further compared the distribution of objects in various classes, including cluster members confirmed by astrometry vs those from the literature, the probable non-members (as defined by PB23), 
and the field population. The analysis is shown in Fig. \ref{fig:index-vs-membership}.
Looking at the variability indices, the new astrometric members from PB23 share the typical variability index distribution of the confirmed literature members and are clearly different from the probable non-members. 
This verifies a low contamination in the sample of astrometric candidates from PB23, as we do not see differences in the indices of the general collection of known members (identified by multiple methods, including spectroscopy, X ray, and the presence of IR excesses) versus those identified by Gaia. 
A further insight in the differences in variability for various types of objects is given in Section \ref{sect:clustering}.

\subsection{Identifying and classifying new variables}
\label{variability-sect}

\begin{figure*}
    \centering
    \includegraphics[width=14cm]{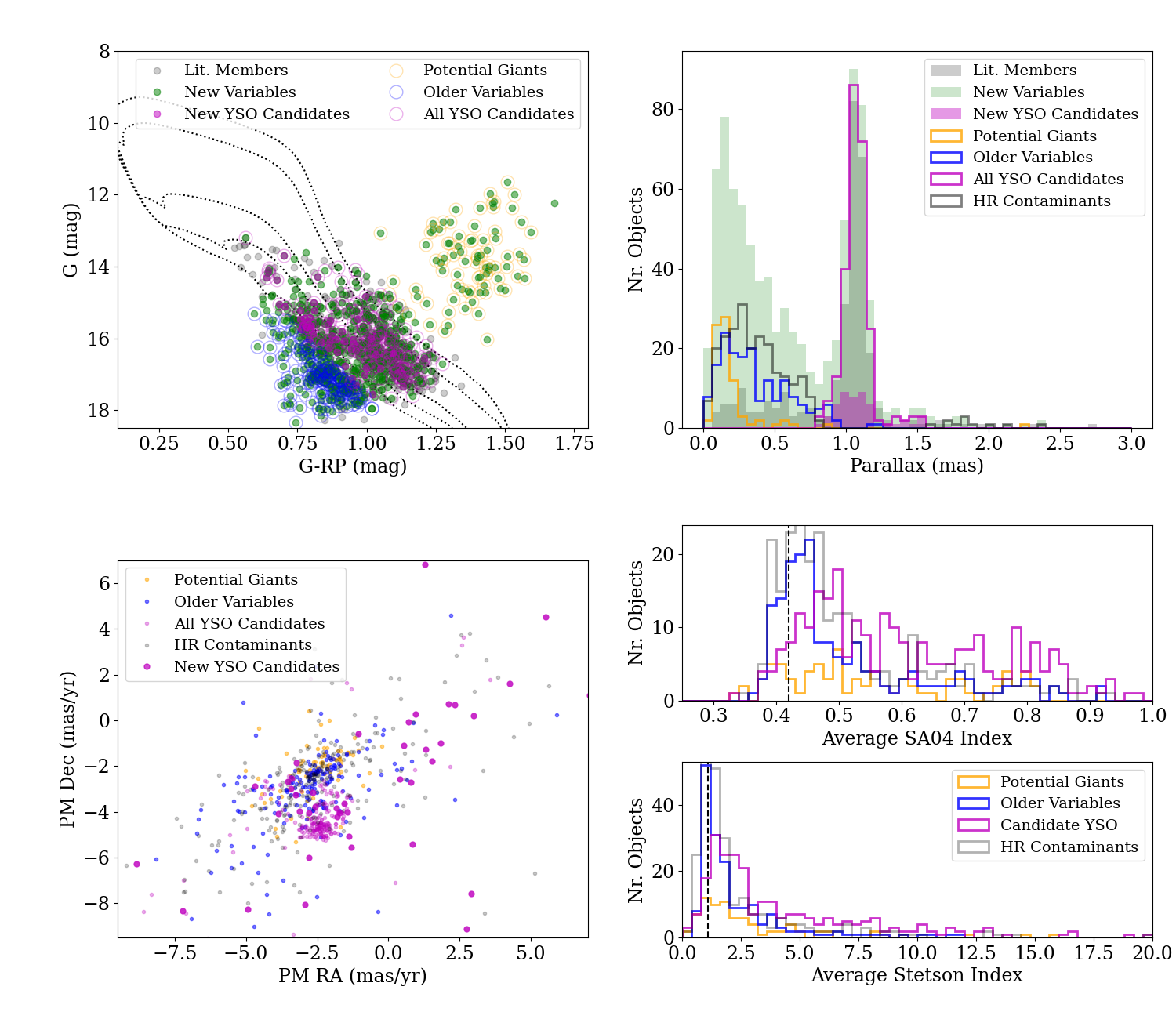}
    \caption{A study of the known and new variable stars. The top left panel show the colour-magnitude diagram for the Gaia filters with the CMD isochrones for ages 0.5, 1, 5, 10 and 20 Myr (corrected by cluster distance and reddening following PB23).
    Members and candidate members derived from the literature are shown in grey, while new variables are displayed in green. Among the new variables, we use the isochrones to select potential 'older' field variables (blue) as well as variable giants (yellow), and use a combination of isochrones and parallax to extract the candidate YSO (magenta). The 'HR Contaminants' have parallax outwith the cluster limits, but fall within the same isochrones than the YSO. 
    The top right panel and bottom left panel show the parallax histogram for the different types of objects, and the proper motions in RA and Dec, respectively. The bottom right smaller panels shows two histograms for the average SA04 and Stetson indices, with the limits for variability shown as dashed lines. To highlight the astrometry of the newly discovered YSO, they are shown with filled histogram in the parallax plot and with larger symbols in the proper motion panel. Note that parallax and proper motions of field variable sources and giants are clearly different and that older variables and HR contaminants have significantly lower variability indices. The proper motion plot is zoomed in so that the detail of the substructure can be appreciated, there are a few high proper motion objects of all kinds beyond the limits we show. The Stetson index is also truncated to 20 for the same reasons. Figure \ref{fig:index_vs_classes} presents further information regarding individual indices within each class.}
    \label{fig:selection-new-variables}
\end{figure*}

After verifying that variability is correlated with youth and membership, we moved onto identifying other young, variable stars in the region. We first constructed histograms of the parallaxes for objects identified as variables via individual indices. 
To define the criteria for variable stars, we started with the most restrictive selection. After eliminating sources that are close to the saturation or noise limits, we also selected those with good Gaia DR3 data. By looking at the distribution of all objects, we observe a clear difference in behaviour for Stetson indices $>$2 and SA04 indices $>$0.5. This difference is further enhanced by requesting objects to be variable in at least two indices of each class and having a minimum of 20 observations in all filters, which allows us to use less stringent limits for similar results, namely Stetson$>$1.1 and SA04$>$0.42.  The H$\alpha$ filter displays a slightly different behaviour and is likely sensitive to higher degrees of variability, but since the difference is small, we are at present treating it in the same way as the rest until more data becomes available. 
We note that the requirement of excellent Gaia DR3 data is failed by a significant number of confirmed members (see PB23). Nevertheless, our aim here is to test the method, for which having reliable Gaia data is an invaluable asset. The requirements about the Gaia data will be relaxed in the future when more data allows us to determine stellar parameters and lightcurves, adding further information to distinguish YSO from other types of variables.

Figure \ref{fig:selection-new-variables} examines the location in the HR diagram, parallaxes, and proper motions for known members compiled from the literature (PB23), as well as new variable stars found by North-PHASE. The initial collection of variable stars was further divided into candidate YSO when positioned between the 0.5 and 20 Myr isochrones in the colour-magnitude diagram \citep[using the CMD isochrones,][]{Bressan2012,Marigo2013} as well as parallax within 0.8-1.6 mas. The isochrone limits are chosen because they contain most of the confirmed Tr~37 members (PB23), and the parallax limits respond to the fact that the background variables and the cluster parallax peak merge at around 0.8 mas, and objects below this limit may have a similar chance to be cluster members or contaminants. We also mark those that are variable but have parallaxes outwith the cluster range, with particular attention to those that fall in the same region of the colour-magnitude diagram, which we label 'HR contaminants'. Objects like these will be the bulk of the contaminants in the literature, since they are harder to distinguish from genuine YSO.  Other variable stars include older systems below the isochrones, and background giants. To explore their variability indices, we consider the average index for each star, including all the available combinations, as well as the standard deviation per object.

Examining the different groups of variable stars reveals that the main sources of contamination are background giants and AGB stars, which are also often variable. Although giants and YSO tend to cover similar index ranges for both the Stetson and the SA04 indices, older variables (below the 20 Myr sequence) and 'HR contaminants' have a different index distribution. As a note, we highlight that PB23 identified some genuine YSO both below the 20 Myr isochrone as well as over the 0.5 Myr one, but also that contamination by field stars is stronger there. While keeping in mind that the selection at this stage is quite restrictive, further study of complete SEDs and accretion signatures will help to find discrepant sources. A double-sided Kolmogorov-Smirnov test (KS test) gives a probability below 10$^{-11}$ for YSO and either older variables or 'HR contaminants' to be drawn from the same index distribution, for both indices. The case for giants is significant for the Stetson average index  (0.2\%), albeit only marginal for SA04 (6\%). Considering the standard deviation of the indices reveals a larger difference between YSO and giants, rejecting the null hypothesis at a level of 10$^{-9}$ for the SA04 indices, and at 0.02\% for the Stetson average index. 

Significant differences also arise when considering individual indices, as shown in Appendix \ref{app-figs}, Fig. \ref{fig:index_vs_classes}. We also see that the second peak in the SA04 indices is more prominent for YSO in the red bands, and that YSO tend in general to have more extreme variability indices. Although processes such as accretion and extinction are expected to produce stronger variations in the bluer bands, the fact that noise is higher in gSDSS for typically-red YSO may play a role in this. HR contaminants show narrower peaks at low index value, especially in the bluer bands. Although at this stage it is too early to define alternative criteria to separate young variables and the rest based on variability alone, the results are promising and suggest that a combination of indices can be used to assign YSO probabilities to objects for which further information and astrometric data are poor or unavailable.

\begin{figure*}
    \centering
\includegraphics[width=18cm]{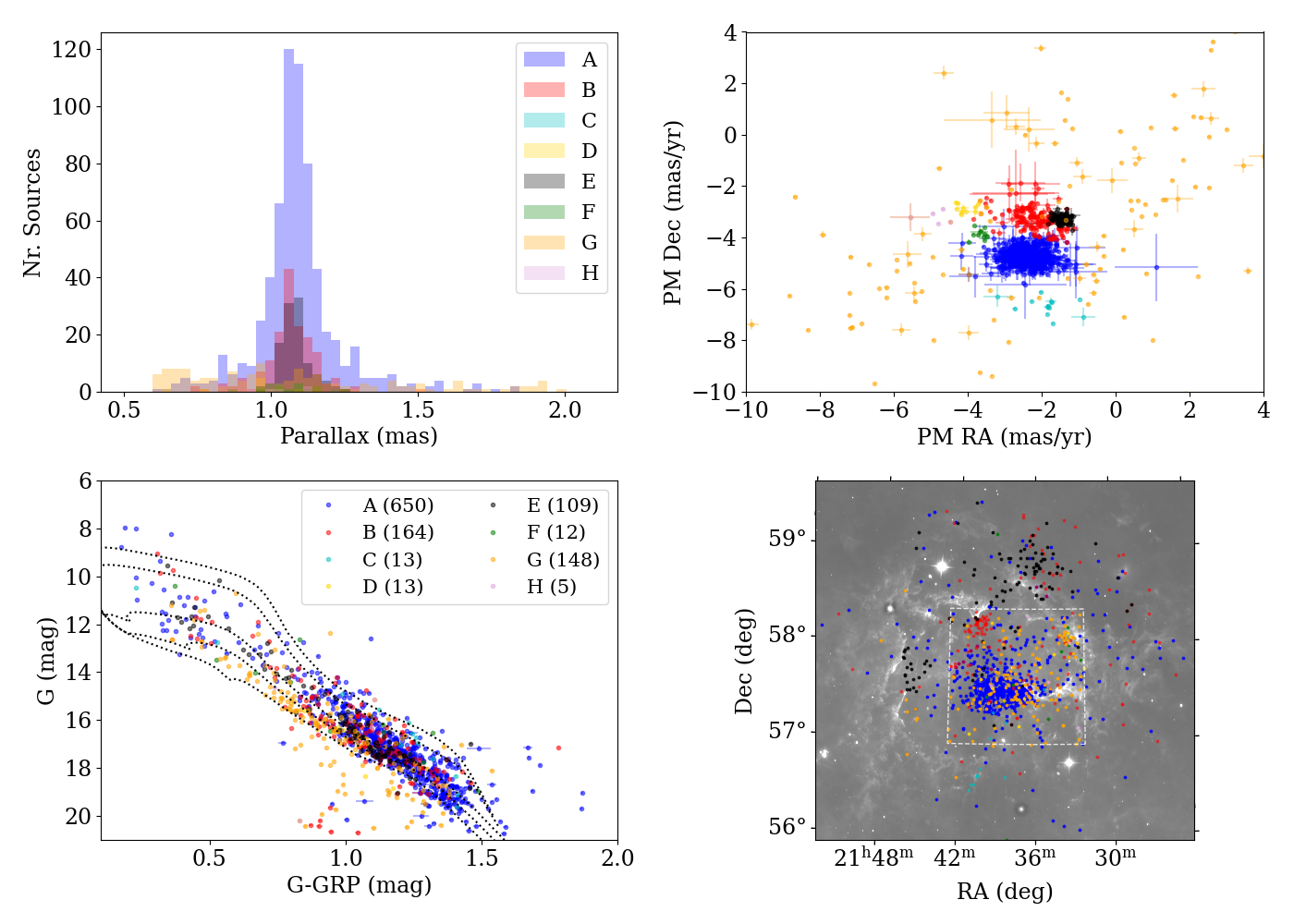}
    \caption{Subclusters in Tr~37, including the members from PB23 and the new variables from North-PHASE. We show the parallax distribution (top left), proper motion (top right), a colour-magnitude diagram (including CMD isochrones for ages 0.5, 1, 5, 10 and 20 Myr corrected by A$_V$=1.5 mag, bottom left), and ICRS coordinates on the WISE mosaic for band W3 (as in PB23), with a white box to mark the North-PHASE FoV (bottom right). The colour scheme follows PB23 for subclusters A-G, adding subcluster H, a new group with low significance. Objects with a probability of at least 40\% of belonging to a subcluster are shown, to include those that have intermediate probabilities (hence the number of objects is different from those in Table \ref{tab:clustering-all}).  Note that the proper motion plot is zoomed in so that the detail of the substructure can be appreciated, there are a few high proper motion objects of all kinds beyond the limits we show.} 
    \label{fig:clustering-all}
\end{figure*}
 
After cleaning the new variable sample using the isochrones, we are left with 50 new members (to add to the 321 literature members that satisfy the strong selection criteria). For a cluster that has been so thoroughly studied as Tr~37, a member increase by 16\% is significant. Among the literature population, 146 stars fall within the variable class. Note that no selection has been done here regarding the parallax of the literature members, hence when clustering is considered (Section \ref{sect:clustering}), we find higher rates of variability for clustered sources. Due to the detailed work of PB23 identifying new members by astrometry, we find that the YSO identified by variability tend to be consistent with the sparse subclusters identified by PB23, or are high proper motion sources, similar to walk-away stars \citep{zhou23} whose understanding is key to test the effects of environment on star formation. The dearth of new members identified within the Tr~37 main cluster proper motions indicates that the member search by PB23 was very efficient, which is also corroborated by the analysis of the variability of the new members (Fig. \ref{fig:selection-new-variables}). 

We also demonstrate the importance of variability to unveil complete YSO populations. In the same way that using Gaia astrometry addressed YSO typically missed by other surveys (e.g. intermediate-mass stars and stars without disks or accretion signatures; PB23), using variability further fills in the gap adding objects that do not share the same astrometry than the bulk of cluster members, such as proper motion outliers.

\subsection{Variability and cluster structure}
\label{sect:clustering}

We analysed whether the subcluster structure of Tr~37 as stated by PB23 changes when considering the new variable YSO, as well as the clustering of objects depending on their variability. As shown in Fig. \ref{fig:selection-new-variables}, the new YSO span a rather broad region in the proper-motion plane, so they add valuable information to investigate distinct star-formation groups and episodes.
We followed the methodology of \citet{roccatagliata20}, based on \citet{Lindegren2000} and also used by PB23. This involves a maximum-likelihood analysis of the tridimensional parameter space including parallax and proper motions in right ascension and declination. Subclusters in different parallax and proper motion ranges were tested against the whole population, and we considered as final subcluster structure the one leading to the higher likelihood. More details on the procedure are given in Appendix \ref{likelihood-app}.

For Tr~37, we did two tests. First, we ran the algorithm from PB23 including the new variable stars identified as cluster members, and then, independently, we used the same technique to examine the distribution of all variable stars in astrometric space, independently of whether they are cluster members. The first part helps to see what the inclusion of variable stars detected over a large FoV can tell us about the global cluster structure. The second part is aimed at improving our understanding of how the astrometry can be used to separate different types of variables and whether YSO and field variables (as grouped by the algorithm) show differences in their variability indices.

\subsubsection{Clustering analysis for all Tr~37 members}

\begin{table*}
    \centering
    \resizebox{\textwidth}{!}{
    \begin{tabular}{ccccccccccl}
    \hline
    Subcluster & $\varpi_{PB23}$ & pm$_{RA,PB23}$ & pm$_{Dec,PB23}$ & $\varpi$ & pm$_{RA}$ & pm$_{Dec}$ & N$_{PB23}$ & N$_{NP}$ & Var. & Comments\\
     & (mas) & (mas/yr) & (mas/yr) & (mas) & (mas/yr) & (mas/yr) & & & \\
     \hline
     \hline
      A  &1.101$\pm$0.130 & -2.432$\pm$0.412 & -4.719$\pm$0.318  & 1.090$\pm$0.144 & -2.382$\pm$0.413 &-4.724$\pm$0.304 & 418(+166) & 625  & 68\% (253) & \\
      B  & 1.098$\pm$0.128 & -2.251$\pm$0.360 & -3.131$\pm$0.252& 1.074$\pm$0.093 & -2.238$\pm$0.433 &-3.220$\pm$0.341 & 33(+80) & 121  & 84\% (51) &\\
      C  &  1.002$\pm$0.035 & -1.835$\pm$0.091 & -6.603$\pm$0.241 & 1.013$\pm$0.041 & -1.884$\pm$0.421 &-6.585$\pm$0.222 & 7 & 10  & 0\% (1) &\\
      D  & 0.963$\pm$0.090 & -3.876$\pm$0.260 & -2.954$\pm$0.301  & 0.973$\pm$0.151 & -4.080$\pm$0.218 &-2.865$\pm$0.148 & 9 & 10  & 86\% (7) &\\
      E  &1.075$\pm$0.060 & -1.374$\pm$0.133 & -3.424$\pm$0.201 & 1.078$\pm$0.046 & -1.433$\pm$0.110 &-3.239$\pm$0.101 & 14(+80) & 87  & 75\% (8) &\\
      F  & 1.092$\pm$0.085 & -3.632$\pm$0.121 & -3.883$\pm$0.143  & 1.077$\pm$0.042 & -3.724$\pm$0.152 &-3.914$\pm$0.131 & 7(+3) & 6  & 100\% (3) &\\
      G  & 0.997$\pm$0.229 & -1.909$\pm$5.159 & -3.124$\pm$4.192  &  1.086$\pm$0.379 & -1.250$\pm$5.648 &  -2.817$\pm$4.496 & 64 & 139 & 55\% (96) & Non-clustered \\
      H   & --  & --  & -- & 1.019$\pm$0.078 & -4.816$\pm$0.109 &-3.128$\pm$0.238 & -- & 3  & 67\% (3) &  Low sign.\\
      \hline
    \end{tabular}
    }
    \caption{Astrometric properties and number of objects for the subclusters identified among the entire YSO population of Tr~37, including stars identified by PB23 and new variables found by North-PHASE. The parallax ($\varpi$) and proper motions are given as the average and standard deviation for each group. The number of objects in each subgroup is restricted to those with a probability $>$80\% (thus not including objects that are consistent with more than one subcluster). For the PB23 results, we give the number of objects first identified among the members, plus the number of new members identified by their astrometric properties using the Mahalanobis distance (given in parenthesis) for the clusters used to identify new astrometric members (A,B,E,F only). Column 'Var.' lists the percentage of objects classified as variable vs the total number of objects with good quality North-PHASE data for each subcluster (in parentheses). }
    \label{tab:clustering-all}
\end{table*}

To study the combination of known and new variable YSO, as in PB23, we require the objects to have good astrometric data and imposed the same parallax cut (0.6-1.6 mas). Since our aim is to see the distribution of stars rather than to improve the subcluster parameters in the search of new members, we did not impose any magnitude or uncertainty restrictions, so that the algorithm was run over a total of 1094 stars, in contrast with the 541 high-quality ones selected by PB23. 
As in PB23, we performed several iterations with various grid parameters (see Appendix \ref{likelihood-app} for details), settling for a 43-point grid $\pm$3$\sigma$ around the subclusters, and using the results from PB23, plus 3 more arbitrary subclusters, as initial trial. Non-significant subclusters are removed by the algorithm. As shown in Fig. \ref{fig:clustering-all} and Table \ref{tab:clustering-all} , we recover all the 6 subclusters (A-F) identified by PB23, including the low-significance groups C and F, plus a similar extended population (G, which likely includes a higher degree of contamination than the rest), and a new subgroup, H, with only 5 members and thus low significance. The astrometric parameters of all the previously-known subclusters are consistent, within the uncertainties. These results highlight the robustness of the cluster structure found by PB23, considering that most of the new YSO have significantly different proper motions compared to the previously known population, while also reinforcing the subclusters that had a lesser representation in their study. %
Since the member list from PB23 includes objects identified by astrometry (their 'new members') from subclusters A, B, E and F, a relatively large increase in objects in those groups is expected independently of the variables. We also note that some stars have large probabilities of belonging to more than one group (e.g. $\sim$40\% for two subclusters). Such objects are not counted as members of any in Table \ref{tab:clustering-all}, which is capped to at least 80\% probability, but they are still considered as cluster members and plotted in Fig. \ref{fig:clustering-all}.

The inclusion of the variables also strengthens the membership of several of the groups surrounding the main cluster, in particular, B, E, and F, and reveals further patterns within the extended population, G. Part of this extended population appears associated with the cloud rim in the north-west (see orange dots in the fourth panel of Fig. \ref{fig:clustering-all}), but we do not find any particular distribution in proper motion for those objects. 
While the proportion of new variable YSO in the different subclusters is similar, we find a larger number of young variables among the extended population, G. Objects with discrepant proper motions are impossible to identify as cluster members via astrometry, a sign that variability studies are a key to fully bring into light cluster members in an independent way, and thus critical to study cluster interactions and the episodic nature of star formation.

\begin{figure*}
    \centering
    \includegraphics[width=18cm]{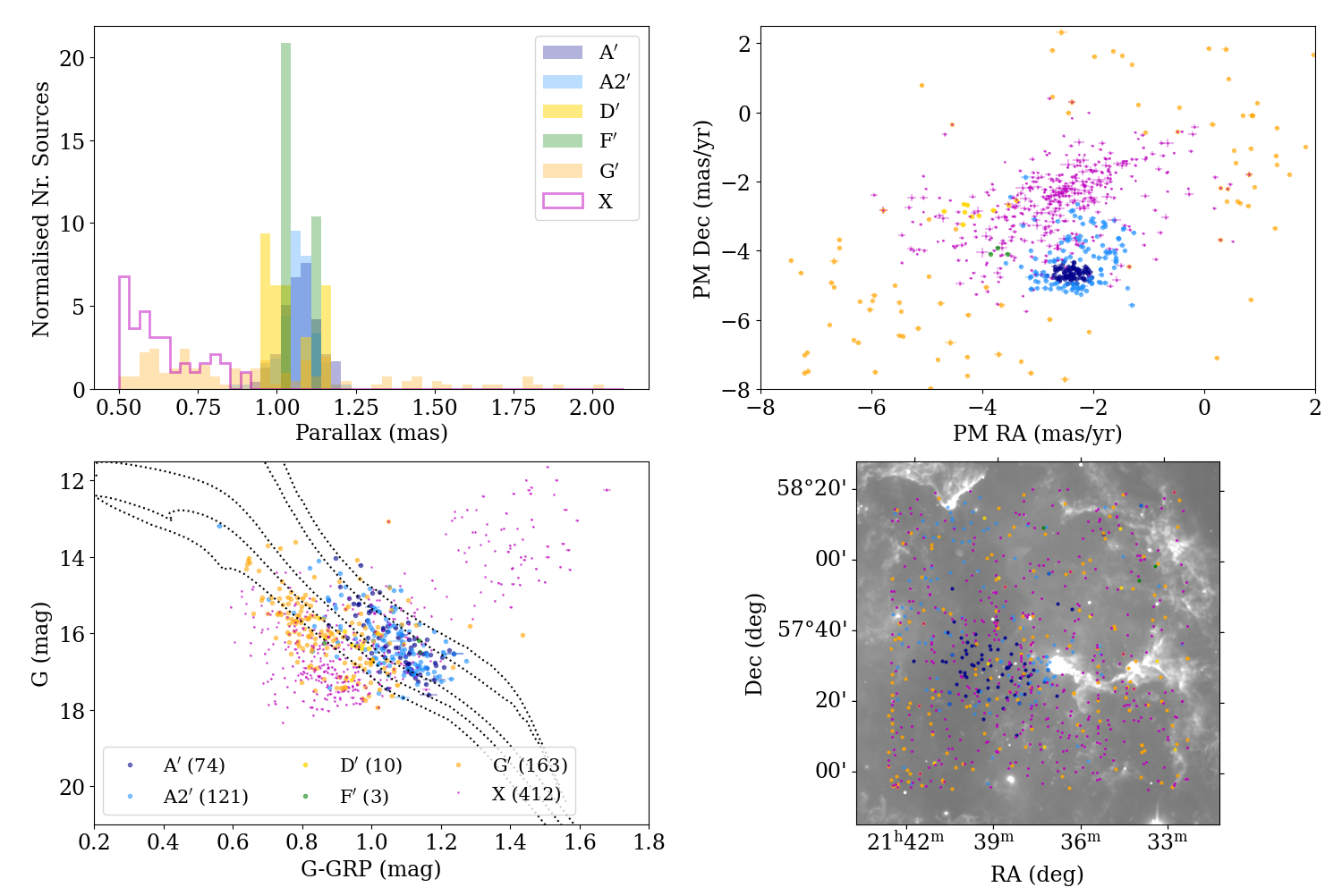}
    \caption{Clustering analysis for all the variable stars in the Tr~37 field (no cluster membership or YSO status required). As in Fig. \ref{fig:clustering-all}, the top left panel shows the parallax distribution, top right panel shows the proper motion distribution (zoomed into the clustered parts), bottom left panel shows the colour-magnitude diagram with isochrones for 0.5, 1, 5, 10 and 20 Myr, and the bottom right panel shows the location of objects over the WISE W3 band image. We recover the main cluster (populations A' and A2', where the later includes some objects from previous populations B and F) and clearly identify background variables.}
    \label{fig:cluster-varia}
\end{figure*}

Table \ref{tab:clustering-all} also contains information on the number of sources that were found to be variable, according to the criteria from Section \ref{variability-sect}. We find a very high percentage of variability among all subclusters with a significant number of members. Typically, at least 2/3 of objects are found to be variable in each subcluster, with the variability rate being higher for sources associated with the globules and bright rims (e.g. B, E), which are also expected to be younger. No difference is found between objects detected by astrometry (PB23 'new members') and those found by other means, a sign that contamination by proper motion and/or parallax interlopers is likely negligible. The extended population, G, also contains a significant, albeit lower, number of variables (55\%), suggesting that, even if contamination may be stronger within this group and some contaminants may be also variable stars (e.g. giants), there is still a significant number of YSO with anomalous proper motions in the cluster.

\subsubsection{Clustering analysis for variable stars}

\begin{figure*}
    \centering
    \includegraphics[width=15cm]{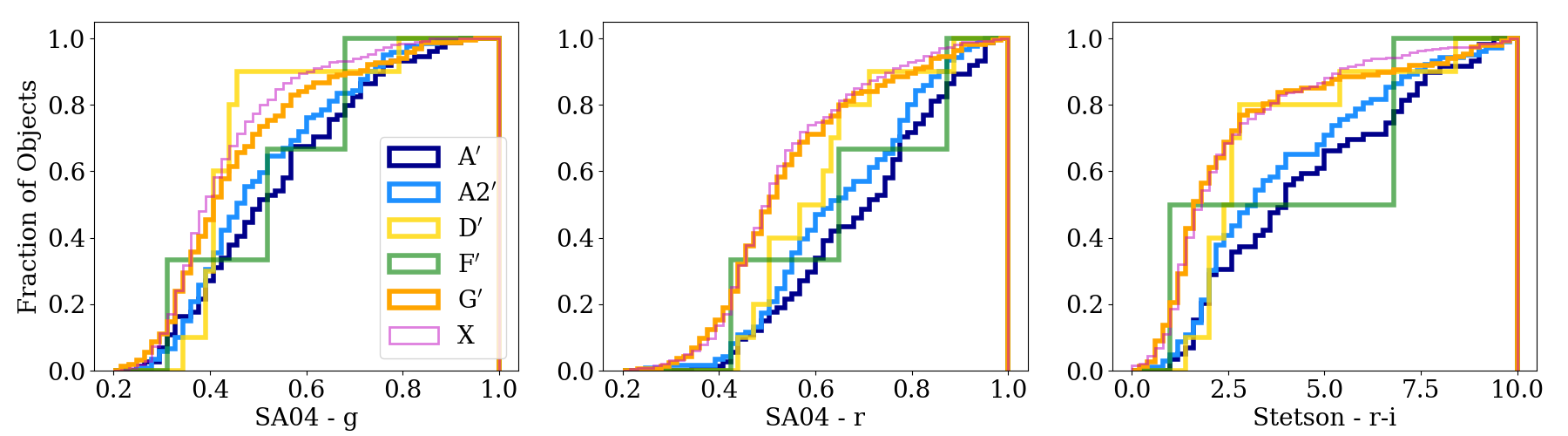}
    \caption{Cumulative distribution of the variability indices for objects classified as variable, according to their subclusters.}
    \label{fig:indexvariahisto}
\end{figure*}

For the second part of the analysis, we repeated the clustering analysis for all the variable stars with good Gaia data (769), without imposing distance or age constraints. One of the objectives here is to determine whether the astrometric clustering can also distinguish different types of variable stars. We ran the maximum likelihood algorithm starting with the clusters defined by PB23 plus two further subclusters to account for the background variables. 
The results are presented in Fig. \ref{fig:cluster-varia} and Table \ref{tab:cluster-varia-table}. We independently recover the selection based on the colour-magnitude diagram from Fig. \ref{fig:selection-new-variables}, identifying the population of variable background giants and older main-sequence stars, as well as the basic structure of the region described by PB23. 
Subcluster E is not found, probably because many of its members are located out of the North-PHASE field, while those of the eastern rim are now set up as a variant of subcluster A (see below). The subclusters with few members (C, H) are also missing, fused with the now-more-extended population G'. The main cluster populations, including Tr~37 (see below) and the surrounding cloud population are identified but the latter, which was previously divided into two main subclusters (E and F in PB23) is now merged together with the group A, which contains the majority of YSO. 

\begin{table*}
    \centering
    \begin{tabular}{lccccl}
        \hline
    Subcluster & $\varpi$ & pm$_{RA}$ & pm$_{Dec}$ & Nr. Sources & Comments\\
     & (mas) & (mas/yr) & (mas/yr) & & \\
     \hline
     \hline
A' & 1.066$\pm$0.055 &  -2.392$\pm$0.122 &  -4.645$\pm$0.109 & 43 & Similar to subcluster A\\
A2' & 1.068$\pm$0.050 &  -2.252$\pm$0.539 &  -4.194$\pm$0.725 & 99 & Contains sources from subclusters A, B and E.\\
D' & 1.035$\pm$0.070 &  -4.150$\pm$0.332 &  -2.818$\pm$0.130 & 8 & Similar to subcluster D\\
F' & 1.078$\pm$0.050 &  -3.632$\pm$0.089 &  -3.997$\pm$0.090 & 2 & Similar to subcluster F\\
G' & 1.049$\pm$0.781 &  -0.655$\pm$6.831 &  -2.372$\pm$6.215 & 145 & Extended/non-clustered \\
X & 0.256$\pm$0.166 &  -2.738$\pm$1.037 &  -2.652$\pm$0.994 & 383 & Background \\
\hline
    \end{tabular}
    \caption{Results of the clustering analysis for all the variable stars. The sublusters are named as variants of the subclusters from PB23, marked with a prime ('). The main cluster, A, splits in two when considering the variable sources only. Variants of subclusters D and F and of the extended population G are also recovered (D', F', G'), together with a pure background group (X). The background population is close to subcluster B in proper motion, but clearly distinct. }
    \label{tab:cluster-varia-table}
\end{table*}

The main subcluster A is identified, albeit divided in two groups, where A' is essentially identical to A and A2' is a variant that contains elements of population B and E as well. Interestingly, the A2' population is not co-located with the A' members, but rather associated with the IC~1396 cloud rims and presenting a larger proper motion spread. The inclusion of the northern globule (globule B in PB23) is logical, since this globule is related to the population B, which is now part of A2', and the same applies to the cloud rim to the east of Tr~37 (former population E). But a new result arises, that population A2' also contains the IC~1396~A globule. IC~1396~A was previously merged with the main Tr~37 cluster, which now appears far more compact. Interestingly, radial velocity differences between IC~1396~A and Tr~37 have been known since long \citep{Patel1995} and further confirmed more recently \citep{sicilia19}, but proper motions were not significantly different when considering the Gaia data for sources known to be physically related to the globule \citep{sicilia19}.
By selecting variable stars only, we may be preferentially choosing the youngest targets, and this suggests a relation between the younger populations in the globules \citep{reach04,sicilia06,getman12} that was not revealed by Gaia data alone nor by individual stellar ages, due to noise. PB23 also identified a difference in age between clustered and non-clustered parts within some of the subclusters, in particular, the subcluster B, although the effect was not evident in others. The youth of the groups is also likely reflected in the high variability percentage found among those groups (Table \ref{tab:clustering-all}). Since the current variability data are limited, this is a result that we will further explore within North-PHASE.

The extended population from PB23 now covers a larger parallax range and probably contains both high proper motion cluster members and field stars, most of which are in the background (subcluster G' in the figure). In addition, the clustering algorithm classifies the background variables in an independent, rather extended population at low parallax (X), clearly distinct from the YSO. The variables at intermediate parallaxes are also very spread in proper motions, as expected from a collection of field stars. From their positions in the colour-magnitude diagrams, they could also include some $\delta$ Scuti variables. A further group with only 3 objects is also identified, but it is not statistically significant and will not be further discussed.

As a final test, we investigated whether the variability properties are different between those 'variable only' subclusters. Figure \ref{fig:indexvariahisto} shows  distribution of the variability indices among the different groups. Objects consistent with cluster members (subclusters A', A2', D', F') are typically more variable and in a larger number of indices. The difference with the background population X is striking, and the extended population G' is also significantly different. A double-sided KS test reveals probabilities under 10$^{-6}$ for all broad-band indices of group A' and X' to be drawn from the same population, and for the SA04 index for H$\alpha$ the probability is 0.01. Comparing A' and G', the probability of both groups to share the same distributions of indices is always below 0.02, and much lower for the Stetson indices (which tend to track the most extreme variables). The difference between A' and G' is more striking than in the case of the members from A and G, probably because the parallax of G' is less well-constrained and thus includes more contaminants. But the results of this analysis confirm significant differences between the variability indices of young variables and variable field stars. These differences offer us a fundamental approach that we will explore in the future to classify objects for which astrometric data are not good or not available.

\subsection{Variability and the presence of protoplanetary disks}
\label{disk-sect}

\begin{table}
   \centering
       \resizebox{\columnwidth}{!}{
    \begin{tabular}{lccccc}
     \hline
    & KS p  &  KS p  & KS p  &  KS p  &  KS p \\
    Index & [3.6]--[4.5] & [5.8]--[8.0]  & [3.6]--[4.5] & [5.8]--[8.0] & ($\dot{M}$) \\
    & (All) & (All) & (Variable) & (Variable) & \\
    \hline \hline
    SA04-g  & 6e-6 & 0.01  & 2e-5  & 8e-3  & 1e-4 \\
    SA04-r  & 4e-7 & 2e-3   & 4e-5  & 0.02  & 7e-3  \\
    SA04-i  & 4e-8 & 1e-4   & 1e-6  & 9e-4  & 1e-3 \\
    SA04-z  & 4e-9 & 2e-4   & 2e-7  & 9e-3  &  7e-4 \\
    SA04-H$\alpha$  & 2e-11 & 5e-5   &  1e-6 & 2e-2  &  5e-5 \\
    Stet. g-r  & 2e-7 & 0.1:  & 3e-5  & 0.05:  & 3e-4 \\
    Stet. g-i  & 3e-7 & 0.08:  & 5e-6  & 0.03  &  2e-3 \\
    Stet. g-z  & 2e-6 & 0.2:  & 3e-5  & 0.04  &  6e-4 \\
    Stet. r-i  & 9e-8 & 3e-4  & 6e-7  & 6e-3  & 1e-4  \\
    Stet. r-z  & 3e-8 & 3e-3  & 4e-6  & 9e-3  &  1e-5 \\
    Stet. i-z  & 5e-10 & 1e-3  & 5e-7  &  1e-3 &  3e-4 \\
    \hline
    \end{tabular}
    }
    \caption{ Results of the KS test comparing the variability indices for YSO with/without disks and accretion signatures. Disks are inferred using Spitzer data for IRAC, with colours [3.6]--[4.5] being sensitive to the innermost disk, while [5.8]--[8.0] can also detect transitional disks with holes and gaps and depleted disks (see Section \ref{disk-sect}). The columns for "All" include all YSO sources from PB23 consistent with the cluster parallax, plus the new variable YSO from North-PHASE. The columns labeled "Variable" include only those known YSO and YSO candidates that are variable. For accreting and non-accreting stars, we use the information from \citet{sicilia10} to determine the accretion status (see Section \ref{sect-accretion}). Non-significant or marginally significant (p$<$0.05) differences are marked with ":".}
   \label{tab:mdot_vs_index}
\end{table}

\begin{table*}
    \centering
    \begin{tabular}{lccccccccccc}
    \hline
   Tentative & Total Nr. & IR excess & IR excess & IR excess & IR excess & SIMBAD & YSO & EB & LPV & Other & Stars \\
   Class  & & (W1-W2) & [3.6]-[4.5] & [5.8]-[8] & [3.6]-[24] &  & & & & & \\
     \hline
     \hline
    YSO & 304 & 59/268 & 40/146 & 74/146 & 37/37 & 220 & 176 & 6 & 1 & 2 & 35 \\
    Giants & 98 & 10/95 & 12/20 & 12/18 & 9/11 &  54 & 12 & - & 38 & 3$^a$ & 1 \\
    Field & 183 & 1/99 & 8/66 & 17/25 & - & 13 & 1 & 6 & - & 4$^b$ & 2 \\
    HRC & 244 & 9/196 & 9/67 & 26/57 & -  & 36 & 17 & 5 & 5 & 7$^c$ & 2 \\
      \hline
    \end{tabular}
    \caption{Summary of the variability types found in Tr~37. For details on the classification scheme, see Section \ref{variability-sect} and Fig. \ref{fig:selection-new-variables}. The excess sources are given relative to the total detected in each band. The classification corresponds to the SIMBAD naming, where we have included all types of YSO together (e.g. TTS, Emission Line, HAeBe, Orion type variable), plus eclipsing binaries (EB) and long-period variables (LPV). At this point, we do not make a distinction between candidates and confirmed types from the SIMBAD database. As 'Other', we include any other object listed as variable but under a different or unspecified category. Under 'Stars' we count any object classified as star by SIMBAD but for which no indication of variability has been reported. $^a$ Including pulsating and SB stars. $^b$ Including one pulsating star. $^c$ Including two RR Lyr and one pulsating star.}
    \label{tab:variability-summary}
\end{table*}

\begin{figure*}
    \centering
    \begin{tabular}{c}
        {\bf {\large{All YSO - Disks based on IRAC [3.6]--[4.5]}}}\\
         \includegraphics[width=14cm]{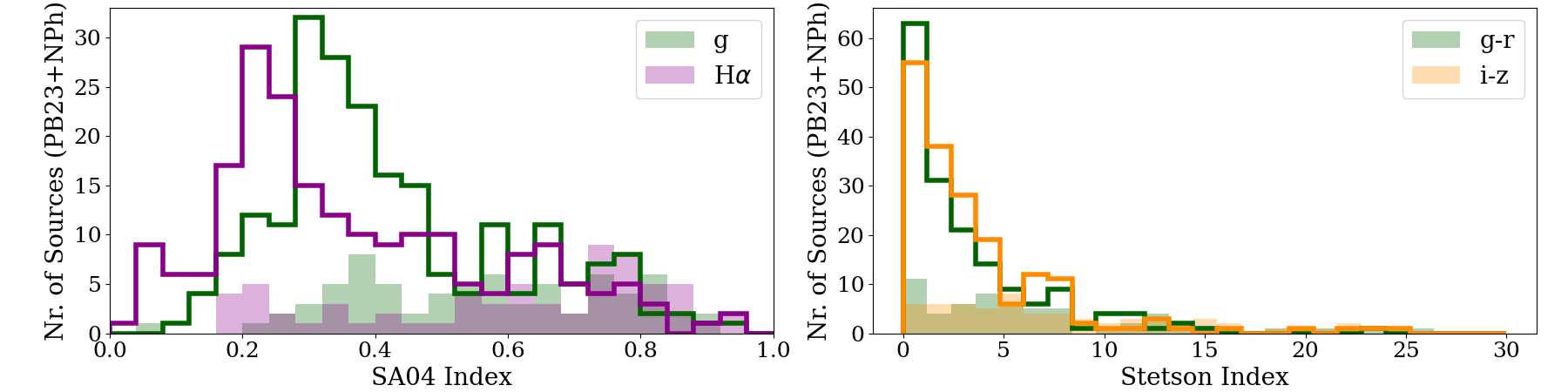}\\
     {\bf {\large{All YSO - Disks based on IRAC [5.8]--[8]}}}\\
              \includegraphics[width=14cm]{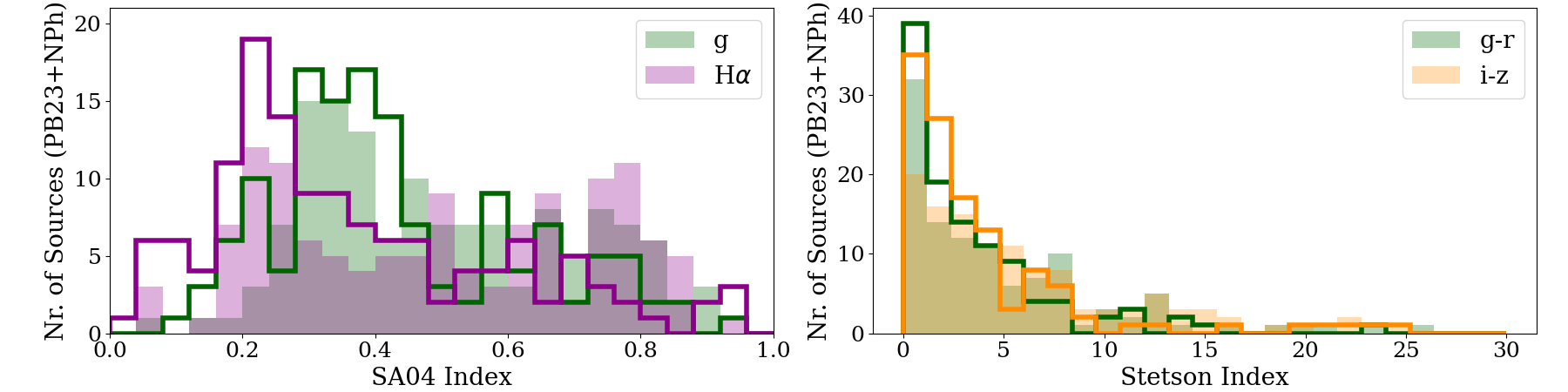}\\
\\
\hline
\\
    {\bf {\large{Variable YSO Only - Disks based on IRAC [3.6]--[4.5]}}}\\
     \includegraphics[width=14cm]{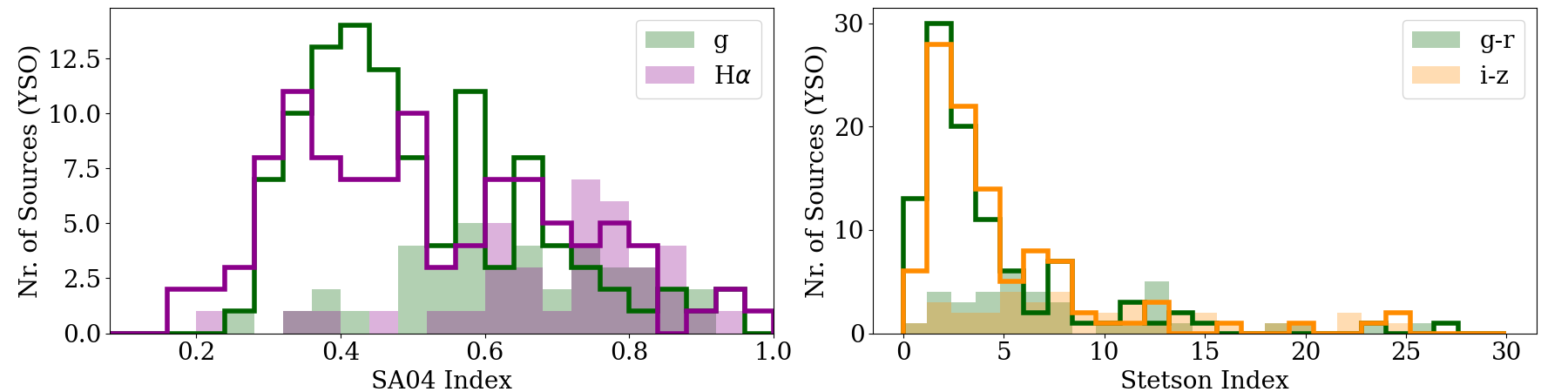} \\
     {\bf {\large{Variable YSO Only - Disks based on IRAC [5.8]--[8]}}}\\
        \includegraphics[width=14cm]{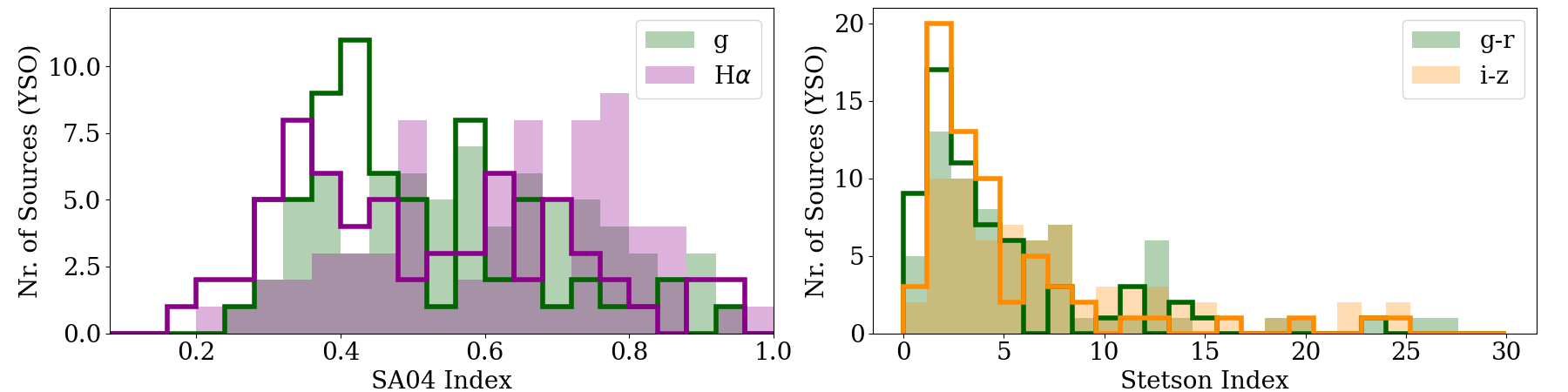}\\
   \end{tabular}
    \caption{ Variability index distribution for Tr~37 stars with (filled histogram) and without (step histogram) IR excess consistent with protoplanetary disks. The top two plots show the results for all the known YSO with Spitzer data, including those listed by PB23 and the new YSO candidates from North-PHASE. The bottom two panels correspond to those of the YSO that are found to be variable, including variable objects from PB23 and newly identified YSO candidates (based on the selection criteria from Section \ref{variability-sect}). Within each set, the Spitzer IRAC colours 3.6$\mu$m-4.5$\mu$m (top) and 5.8$\mu$m-8$\mu$m (bottom) are used to characterise the excess. Note that the Stetson index axis is truncated to value 30 so that the details at low indices are visible; there are very few objects beyond this limit and the KS tests (Table \ref{tab:mdot_vs_index}) include them all.}
    \label{fig:indexvsexcess}
\end{figure*}

Since protoplanetary disks are one of the key features of YSO related to many variability traits \citep[e.g.][]{herbst94},  we looked  at the variability of YSO with and without disks (inferred from their IR excesses) as well as at IR excesses among sources identified as variable by North-PHASE. The first part reveals the variability properties associated with the disks, tracking index performance regarding disk-related processes and what could be lost when new YSO are found by variability. The second part helps us to put in context all variables identified by North-PHASE.

\subsubsection{Variability for disked and diskless sources}
\label{sect-disksYSO}

The most detailed disks studies of Tr~37 were done using Spitzer \citep{sicilia06,sicilia13}, so we used Spitzer data to explore disks vs variability for all the known members within the parallax limits imposed by PB23.  Tr~37 was observed with Spitzer using both Infrared Array Camera \citep[IRAC,][]{fazio04}, with wavelengths 3.6 to 8$\mu$m, and the Multiband Imaging Photometer \citep[MIPS,][]{rieke04} at 24$\mu$m. The cluster is not entirely covered, but we find 284 sources that have both Spitzer observations and North-PHASE data, using a 2" radius to match (PB23).

Disks were identified by requesting IRAC [3.6]--[4.5]$>$0.2 mag, versus [5.8]--[8.0]$>$0.2. The first criterion identifies objects with a near-IR excess, but misses transitional and depleted disks. The second criterion is better suited at finding evolved disks, although it is also more susceptible to contamination. Using [3.6]--[4.5], we find a disk fraction of 25\%, while this raises to 48\% with [5.8]--[8.0]. These results are consistent with \citet{sicilia06}. We also find that 81\% of disks with  [3.6]--[4.5] excess are variable (58/72), compared with 58\% (123/212) of those without disks. When [5.8]--[8.0] is considered, the fraction of variables changes to 70\% (disked, 95/135) vs 60\% (diskless, 82/136). The first conclusion is that, even though we are still at the beginning of the survey, variability can detect about 2/3 of YSO both with and without disks. The second, that disks with excesses in the near-IR bands are more variable. This probably results from those disks being less evolved \citep{sicilia06} and having higher accretion rates \citep{sicilia10}, making them more prone to variability caused by hot spots, accretion, and circumstellar extinction events.

We also examined potential differences between the variability indices per among each group via histograms and double-sided KS tests. Figure \ref{fig:indexvsexcess} shows some examples of the variability index distribution for objects with and without disks as defined above. Although disked sources tend to be more variable (especially, if they have near-IR excess, as mentioned before), there are diskless ones with similarly high variability indices, and most YSO are variables. Table \ref{tab:mdot_vs_index} shows that the differences in index distribution are statistically-significant for all the SA04 indices (specially, for H$\alpha$) as well as for most of the Stetson indices. These differences also highlight the role of various types of indices to obtain a complete view of variability in YSO.

\subsubsection{IR excess among variable stars}

For the second part, we compared the 894 variable sources with the Wide-field Infrared Survey Explorer \citep[WISE][]{wright10} as well as with Spitzer. A 2" radius finds 707 counterparts in WISE, while 309 stars were detected in at least two Spitzer bands. The advantage of WISE is that it covers the entire region; the advantage of Spitzer is its better spatial resolution, making a clear difference in crowded areas and around nebulosity and allowing to detect disks in more advanced stages of evolution. As shown by PB23, WISE W3 (12~$\mu$m) and W4 (22~$\mu$m) have too much contamination in Tr~37, so we can only identify disks using W1 (3.4~$\mu$m) and W2 (4.6~$\mu$m). This means that depleted and transition disks with holes, gaps and/or small excesses may be missed, so disk fractions are a lower limit. 
We then examined the IR colours for the variable stars, subdividing them as in Section \ref{variability-sect} and Fig. \ref{fig:selection-new-variables}.  We focused on the 4 groups of likely YSO, candidate giants, older sources, and 'HR contaminants', and classify as excess sources those with colours $>$0.2 mag (PB23). Detailed figures are shown in Appendix \ref{app-figs}.

We find a substantial number of YSO variable sources with disks, although some diskless sources have variability indices comparable to disked ones, strengthening the result from Section \ref{sect-disksYSO} that variability detects YSO with and without disks. For WISE, 63/291 sources have disks as seen in W1-W2 (disk fraction among variables 22\%). For Spitzer, 42/149 have[3.6]--[4.5] excesses (28\%), 76/149 have [5.8]--[8.0] excesses (51\%). Since MIPS did not reach the photospheric levels of Tr~37 objects later than A-type \citep{sicilia06}, all sources found in this band have excesses. Although large variability indices are more frequent among disked sources, there are diskless YSO with similarly large indices, even though the distribution of indices is clearly different among both groups. The disk fractions among variable YSO candidates are similar to the general disk fractions in the known YSO population, which is a further sign of variability efficiently picking both disked and diskless YSO. Figure \ref{fig:indexvsexcess} compares some of the variability indices for disked and diskless sources, their differences highlighted by the KS tests in Table \ref{tab:mdot_vs_index}. We find that the distributions of all variability indices for objects with and without disks are significantly different, a further point that North-PHASE alone can add valuable information about the evolutionary status of YSO.

In contrast, there are essentially no 'older variables' nor 'HR contaminants' with excesses, but they are found for a few 'candidate giants', including 24$\mu$m excess.
Although IR excess among giant stars are common \citep[e.g.][]{groenewegen12},
we checked whether some could be very young stars or protostars, which often share the same region of the colour-magnitude diagram. Among 'candidate giants' with [3.6]--[4.5]$>$0.75 mag (the range of flat disks and protostars), we find several known TTS with massive disks, including 14-141, 11-2322, and 13-669 \citep[the first two belong to the IC1396A globule population,][]{sicilia06}, two sources with full disks observed with Herschel \citep[21392541+5733202 and 21393104+5747140;][]{sicilia15}, a young X-ray source \citep[21384282+5728547;][]{mercer09}, and two emission line stars \citep[119 and 420 from][]{nakano12}. There is a further object with parallax 1.3 mas, which is likely a newly discovered YSO (at 21:37:43.87 +57:15:36.6) and 3 background variable sources with low parallax, which are likely variable giants. The number of YSO in this group is a sign that the isochrone cut is probably too restrictive, due to the compromise to limit contamination.

Among the 8 'candidate giants' with MIPS 24$\mu$m excess, we recover source 420 from \citet{nakano12}, but the rest are background sources. Three of them are listed as long-period variables by SIMBAD \citep{heinze18,gaia22}, one is a potential binary foreground star with parallax 2.2~mas, and the rest are newly discovered background variables. This demonstrates that North-PHASE is also able to detect other variables, including long-period ones, a sign of the returns that it can provide to the community beyond star formation. 

The conclusions are that our YSO selection criteria are very powerful to find real young stars, although some of them are lost to the 'giant candidates' and 'field stars' groups. The risk of missing sources to the 'field' is higher for dippers and objects with UXor variability (which are fainter and bluer in eclipse), but classification will improve as more variability data is collected, and considering the variability amplitude and lightcurves. YSO that fall over the isochrone into the 'giants' region tend to be very young and embedded and have massive disks and/or envelopes, so they can be further identified. The differences between the variability types within each class suggested in Section \ref{variability-sect} hint that future study of this aspect will also aid the classification. The details are given in Table \ref{tab:variability-summary}, which includes the summary of all the variable stars with IR counterparts as well as those that have a SIMBAD record about its class.

\subsection{Detecting accretion and activity with North-PHASE}
\label{sect-accretion}

Accretion and chromospheric activity are key causes of YSO variability \citep[e.g.][]{fischer22}. At the point of writing this introductory paper, stellar masses and luminosities, required to measure accretion, are not available yet. The number of uJAVA observations, which is key to determine absolute accretion rates \citep{gullbring98}, is still very small. Nevertheless, we can already explore whether the existing H$\alpha$ and variability data correlate with the known accretion rates to test future capabilities.

For Tr~37, accretion rates or their limits have been measured from U-band photometry and spectroscopy for about a hundred objects \citep{sicilia10}, and via H$\alpha$ photometry for about 150 sources \citep{barentsen11}, some of which are in common. Further H$\alpha$ measurements from spectroscopy are available for an even larger sample, as it has played an important role in identifying a substantial number of members \citep{sicilia04,sicilia05cepob2,sicilia06,sicilia13,nakano12}. 
There are many different proxies to derive accretion rates \citep[e.g.][]{gullbring98,fang09,alcala_x-shooter_2014}, not entirely deprived of systematic differences. Before calibrating the North-PHASE data, it is important to explore the agreement within existing measurements, which is not always evident (see Appendix \ref{accretion-app}) and will need to be investigated in the future.

\begin{figure*}
    \centering
    \includegraphics[width=14cm]{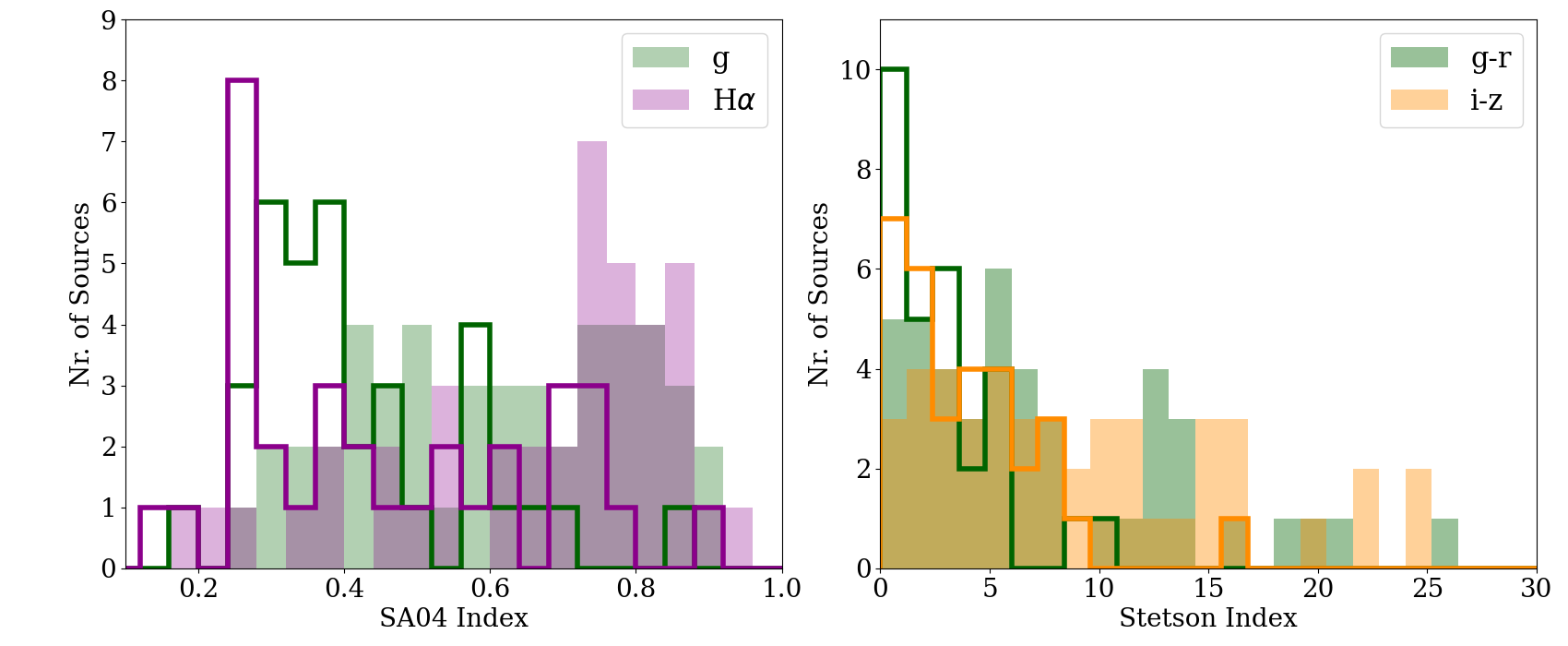}
    \caption{A comparison of the variability index distribution for stars with accretion (filled histogram) vs no accretion (step histogram), using the information on the accretion status from \citet{sicilia10}. The left panel shows some examples of SA04 indices, while the right one shows examples with the Stetson indices. Note that the Stetson index x axis is truncated to value 30 so that the details at low indices are visible; there are very few objects beyond this limit and the KS tests (Table \ref{tab:mdot_vs_index}) include them all.}
    \label{fig:index_accretors_histo}
\end{figure*}

We first compared the variability indices for stars with and without known accretion. For this, we used the data from \citet{sicilia10}, finding 82 cross-matches with the North-PHASE lists. The accretion data includes detections, upper limits, and objects for which accretion was ruled out using high-resolution H$\alpha$ spectroscopy. Although correlations between indices and accretion rates are at most marginal (only indices involving gSDSS band seem to show some correlation, with Spearman r around 0.4 and probability p around 0.01), we observe that the indices for sources with and without accretion show clear differences (see Fig. \ref{fig:index_accretors_histo}). Table \ref{tab:mdot_vs_index}, containing the results of a double-sided KS test between the indices of non-accreting (35) and accreting (47) sources, demonstrates its significance. The differences show that the magnitude of the variability index can reveal information about the accretion status of the sources. Although the trend was already seen when checking the indices of disked sources (Fig. \ref{fig:indexvsexcess}), we now clearly see that the second peak of the SA04 indices is strongly biased towards accreting sources, which explains its correlations with YSO previously mentioned.

The results are in line with the previous discussions of the SA04 and Stetson indices.
As expected, the H$\alpha$ band shows the strongest differences among the SA04 indices.  Although switching from an accreting to a non-accreting status is very rare \citep{sicilia10,sicilia20-j1604}, accretion rates for individual stars are typically variable by factors of few to tens, which may dilute the correlations between indices and accretion rate. It is thus of paramount importance to explore the results of the u-JAVA filter in the future, to be able to quantify accretion at the same time than variability.

\begin{figure}
    \centering
    \includegraphics[width=8cm]{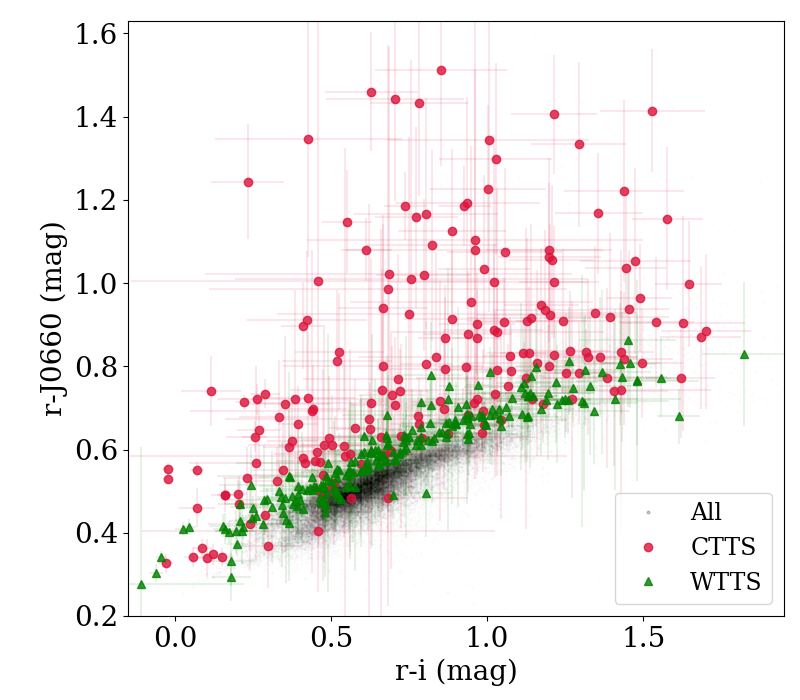}
    \caption{Colour-colour diagram involving average rSDSS, iSDSS, and J0660 colours, to explore H$\alpha$ excesses. Small grey symbols show the field objects, red dots show known CTTS, and green triangles display known WTTS, classified based on existing spectroscopic and photometric H$\alpha$ EW. The'error bars' are not the photometric errors (which are much smaller), but the standard deviation of the magnitudes. The diagram shows how the r-J0660 colour distinguishes the CTTS from the WTTS and from the general population. Note that CTTS and WTTS have been corrected for the general cluster extinction, but the rest has not been modified. }
    \label{fig:Hacolour}
\end{figure}

We also used the J0660 filter to detect accreting and active stars via a colour-colour diagram, following \citet{barentsen11}. We considered the average rSDSS, iSDSS and J0660 magnitudes and standard deviations, exploring both field stars and YSO with estimated H$\alpha$ equivalent width \citep[EW; from][]{sicilia05cepob2,sicilia06hecto,sicilia13,barentsen11,nakano12}. For those, we classified them according to their published H$\alpha$ EW as classical T Tauri stars (CTTS) or weak-lined T Tauri stars (WTTS) using a simple threshold of 10\AA\ (noting that this threshold may overestimate the CTTS in the lower mass range, while underestimating them towards the higher mass range). Only objects with at least 5 observations in all the involved filters were considered. The results are shown in Fig. \ref{fig:Hacolour}. The young CTTS and WTTS stand out from the field stars and from each other, which validates the use of the North-PHASE photometry to distinguish YSO as well as accreting and non-accreting (in case of WTTS and YSO, chromospherically active) stars.

Although a proper estimate of the H$\alpha$ EW requires model photospheres \citep[e.g.][]{barentsen11}, which is beyond our initial capabilities, we demonstrate that the process is feasible using the North-PHASE data. As before, we note that we are comparing non-simultaneous measurements, which will increase the spread. A rough estimate of the flux in the H$\alpha$ line can be obtained from the J0660 flux, taking a scaled version of the rSDSS flux as the continuum. For this first attempt, we used the data acquired on the reference night. We used the filter zero point calibrations and FWHM from the Spanish Virtual Observatory \footnote{\href{http://svo2.cab.inta-csic.es/svo/theory/fps3/}{http://svo2.cab.inta-csic.es/svo/theory/fps3/}, see filters for the Observatorio Astron\'{o}mico de Javalambre (OAJ).} for the conversion, including the average cluster extinction from PB23 (A$_V$=1.4 mag). 

The resulting H$\alpha$ EW was compared to EW measurements from spectra \citep{sicilia05cepob2,sicilia06,sicilia13,nakano12} and IPHAS photometry \citep{barentsen11}. To check if different calibrations could be an issue here, we made two tests: using the Pan-STARRS calibration for our rSDSS data that we are adopting for the survey (see Section \ref{sect:fluxcal}), or using IPHAS to calibrate rSDSS (note that J0660 is always calibrated using IPHAS), but the results, despite systematics, are comparable. The procedure of estimating 'rough' EW is particularly inaccurate for late-type stars where the red part of the spectrum is very far from flat, which triggers a general offset in the measured EW, but we find a clear correlation between available EW measurements (Spearman r $\sim$ 0.3 for spectroscopic results, or 0.5 comparing to photometric values, with essentially negligible probability that the results are not correlated). This shows that the North-PHASE J0660 photometric measurements can be used to estimate the strength of the H$\alpha$ line, as well as to distinguish stars with and without emission, which adds to our criteria for the identification of YSO, even if the systematic differences will require an improved calibration (see Appendix \ref{accretion-app}).

\subsection{Detecting rotation and classifying light curves}

At present, the North-PHASE data on Tr~37 is not complete enough to independently derive rotational periods. Nevertheless, we did some brief tests using sources for which information on periods and light curves was available from the Transiting Exoplanet Survey Satellite (TESS), to assess the feasibility of using North-PHASE data to detect rotational modulations and distinguish the causes of variability in objects in a wide range of magnitudes. 

We focused on a few well-known objects classified as different types of variables using TESS data (see Kahar et al. in prep., Schlueter et al. in prep.) from the sectors available for Tr~37 (16, 17, 56 and 57). The data were extracted using {\sc Python TESSCUT} and a 3 pixel window. For some stars, the presence of similar-magnitude, nearby ($\sim$20") objects casts some doubts regarding the possible contamination, but we marked them accordingly and use them as a test of what North-PHASE can detect. The risk is higher for fainter targets. 

The periods were derived from the TESS data using a Lomb-Scargle periodogram \citep{vanderplas18} followed by phase-folding to examine the resulting light curves. 
Note that, since most of the stars have multiple variability causes, ranging from rotationally modulated (hot and cold spots) to quasi-periodic (dips) and irregular (outburst), photometric periods of young stars are often not detectable in all datasets or during all epochs \citep{sicilia23}. The information from the TESS analysis is then used for a consistency test of the North-PHASE data. Phase-folding of the North-PHASE data allows to check if it agrees with the known TESS period and infer other potential periods (particularly helped by the densely-sampled interval around JD=2460039-47, although it will need to be confirmed with further data), while examining the lightcurves allows to compare the typical features detected by TESS (e.g. dips, bursts, sinusoidal modulations and irregular activity).

The selected sources are listed in Table \ref{tab:period-check}, {\b and a few examples are shown in Fig. \ref{fig:rota-example}. The naming convention follows \citet{sicilia05cepob2}. Their magnitude range spans gSDSS$\sim$13-19 mag, which ranges from non-saturated data to the current noise limit.
We find that all sources are labeled as variables in most filters, considering the SA04 indices although, as expected, those faint and red are often noisy in g, preventing the detection of variability at lower rates. 
For 13-277, also known as GM Cep and known to have variable accretion and dipper events \citep{sicilia08gmcep,semkov12}, we do not observe rotational modulation, but find a dip. Object 11-2131 is inconclusive, and we also fail to detect the 1d period of 72-875 due to sampling or to the effect of nearby objects in the TESS datasets, but the remaining objects all display different levels of modulation. As seen in Fig. \ref{fig:rota-example}, the levels of the rotational modulation observed have peak-to-peak changes typically below 0.2 mag. 
As expected, the North-PHASE data sampling in intervals of 1-3 days allows to verify the typical rotational periods in the range of 1-10 days \citep{bouvier_coyotes_1995}.
The conclusion from this limited experiment is that North-PHASE can detect stellar rotation for a large number of young stars, and distinguish between the different variability types in objects that may not be dominated by rotational modulation, or may present quasi-periodic variations. 

\begin{figure*}
    \centering
    \begin{tabular}{cc}
        {\Large {\bf 13-277}} & {\Large {\bf 11-2037}}\\
    \includegraphics[trim=2.0cm 0 0.5cm 0, width=9cm]{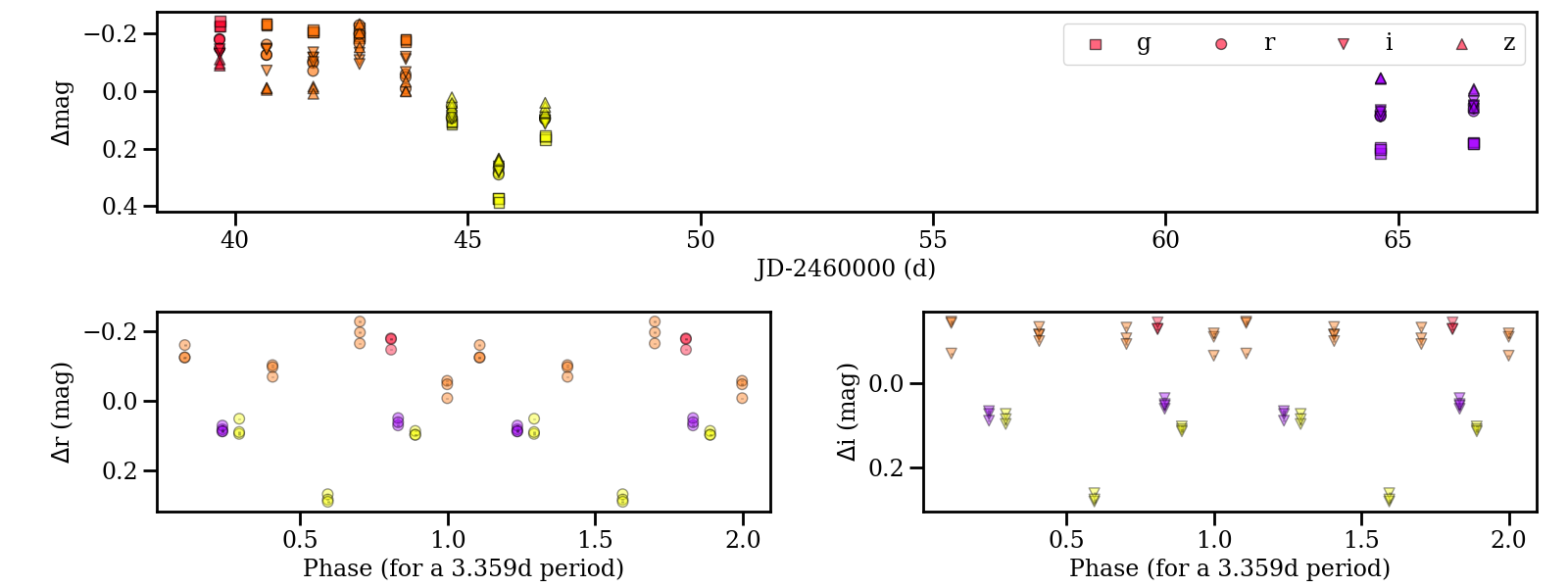} &
    \includegraphics[trim=2.0cm 0 0.5cm 0, width=9cm]{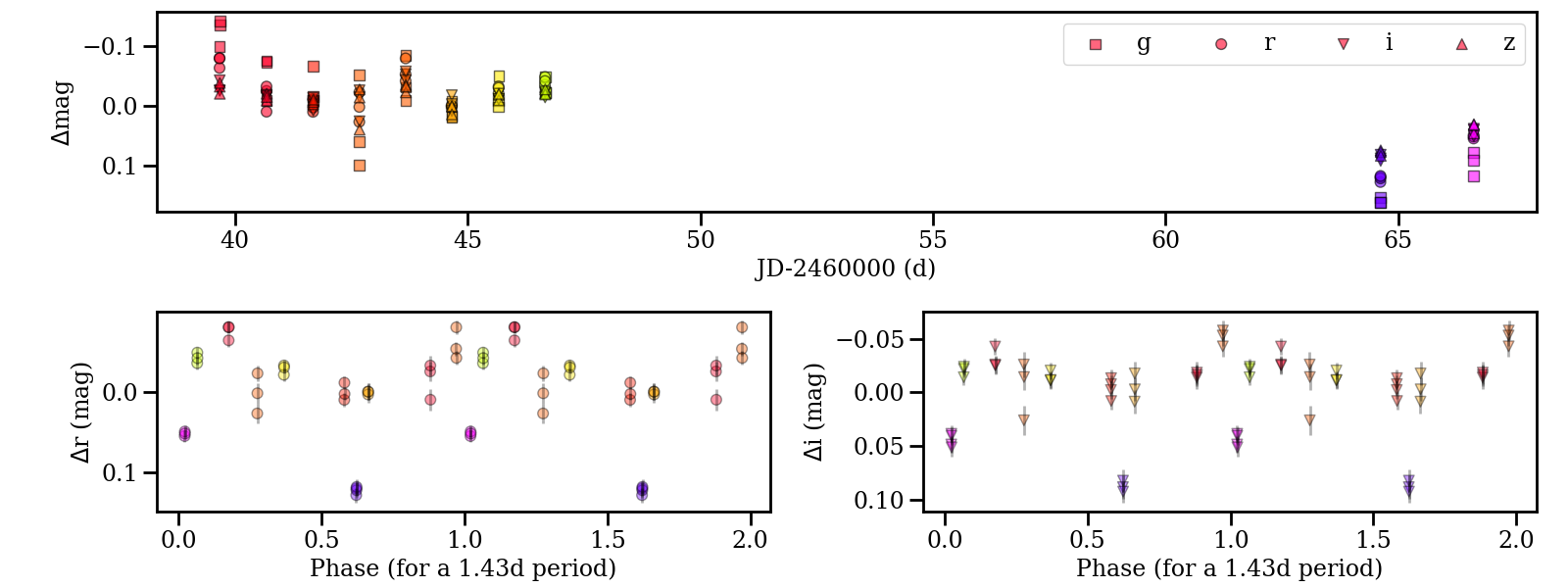}\\
    {\Large {\bf 73-71}} & {\Large {\bf 73-758}}\\
      \includegraphics[trim=2.0cm 0 0.5cm 0, width=9cm]{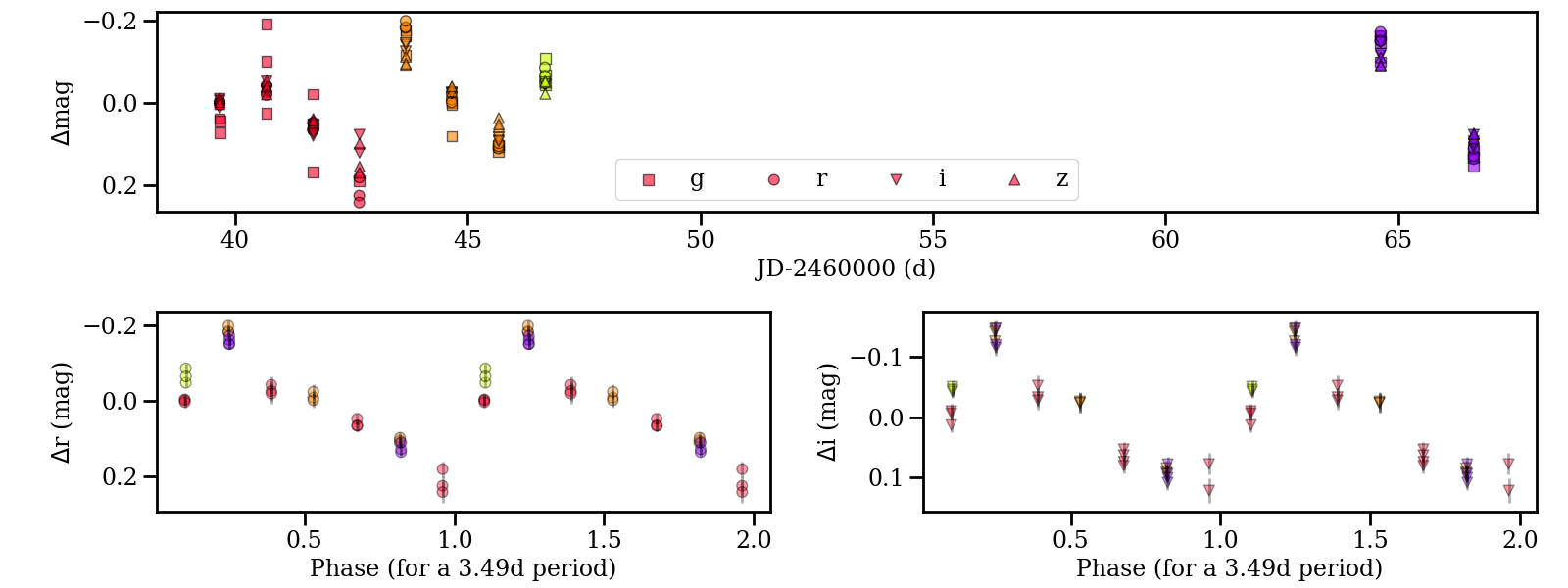} &
        \includegraphics[trim=2.0cm 0 0.5cm 0, width=9cm]{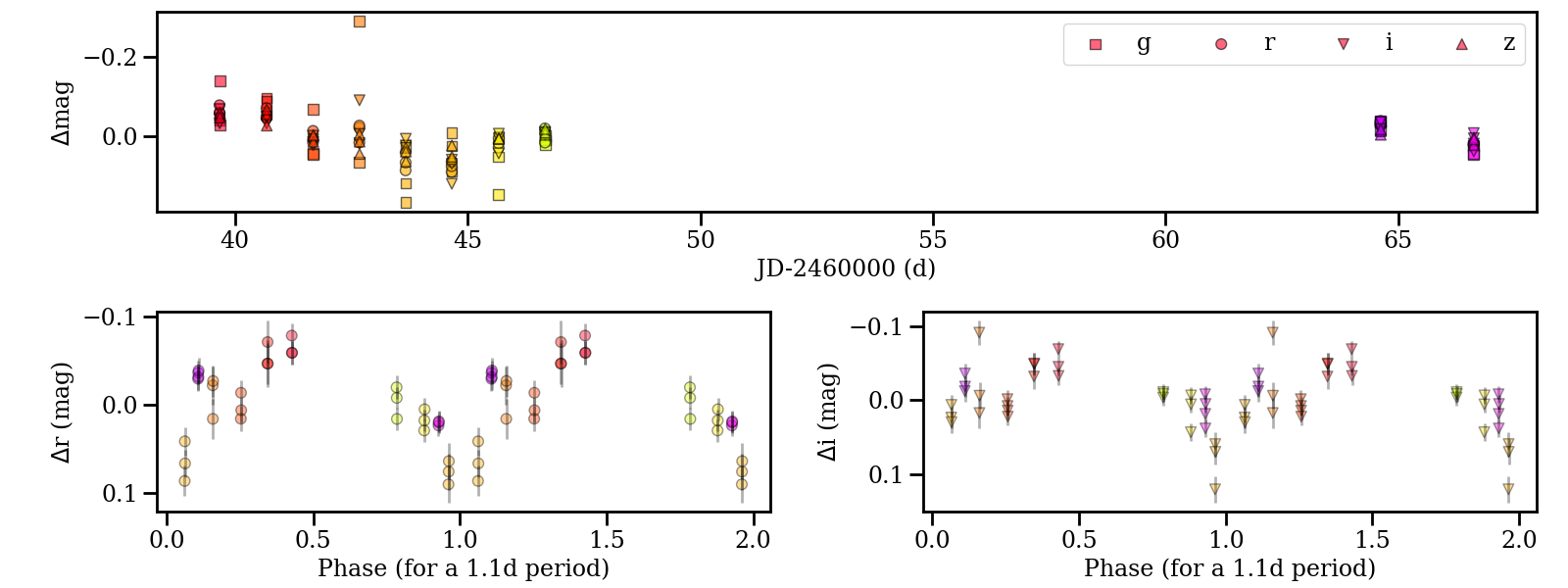} \\
    \end{tabular}
    \caption{Some examples of rotational modulation observed by North-PHASE in Tr~37. For each target, the top panel shows the complete light curve relative to the average magnitude in each filter (note that dates are colour-coded, while filters are marked by different symbols). The two bottom panels show the phase-folded light curve for the best period for the rSDSS and iSDSS filters. }
    \label{fig:rota-example}
\end{figure*}

\begin{table*}
    \centering
    \begin{tabular}{lcccccc}
    \hline
Name & gSDSS & P$_{TESS}$ & P$_{NP}$ & Variability & Type & Type \\
 & (mag) & (d) & (d)& (North-PHASE) & (TESS) &  (North-PHASE)\\
    \hline\hline 
13-277  & 13.58$\pm$0.22  & 3.59 & --  & grizH$\alpha$/All St & Rot./Dip. &  Irr./Dip.\\
12-1091 &  16.89$\pm$0.10 & 7.5: & 3.6: &  grizH$\alpha$/All St &  Rot.:/Dip. & Rot./Dip. \\
11-2037  & 16.91$\pm$0.08  & 1.43  & 1.43 & grizH$\alpha$/All St & Rot. & Rot./Unc. \\
14-125  & 17.31$\pm$0.14  & 9.9:$^*$ & 9.5: & griz/All St & Rot./Irr.: & Rot. \\
73-71  & 17.90$\pm$0.11  & 3.49 & 3.49 & griz/All St & Rot. &  Rot. \\
14-183  & 17.65$\pm$0.16  & 6.44:$^*$ & 6.4 & grizH$\alpha$/All St    & Rot./Dip./Irr.  & Rot./Dip.: \\
73-758  & 18.01$\pm$0.09  & 1.10$^*$ & 1.1 & rizH$\alpha$/All St  & Rot./Bur. & Rot. \\
13-1048  & 18.01$\pm$0.27  & 1.8:$^*$ & 9.9: & grizH$\alpha$/All St    & Irr. &  Rot./Dip. \\
11-2131  & 18.36$\pm$0.25  & -- & 5.58 &grizH$\alpha$/All St & -- &  Unc./Rot.: \\
21-998  & 18.93$\pm$0.46  & 9.69$^*$ &  9.7 & grizH$\alpha$/All St   &  Rot.:/Dip./Irr. & Rot.  \\
72-875  & 19.14$\pm$0.21  & 1.04:$^*$ & 4.95: & grizH$\alpha$/All St  &  Rot. & Rot.:\\
92-393  & 19.17$\pm$0.17  & 1.40$^*$ & 3.3: & rizH$\alpha$/Red St & Rot.: &  Rot.: \\
13-1238 &  19.23$\pm$0.21 & 7.99: & 7.1: & grizH$\alpha$/All St   &  Unc. &  Rot. \\
\hline
    \end{tabular}
    \caption{Comparing the variability detected by TESS and North-PHASE. We provide the calibrated, average North-PHASE gSDSS magnitude and standard deviation (a measurement of the spread). We show the best estimated period from TESS (from periodograms) and North-PHASE (as visually inferred from phase-folding of the limited data), if available. Note that, at present, the North-PHASE proposed periods are based on consistency with TESS and visual inspection of the phase-folded curves, being merely indicative. Sources with TESS data that may be compromised to some extent due to others nearby are marked with $^*$. All sources are found to be variable in most filters, with H$\alpha$ being compromised when close to nebulosity (e.g., for 14-125 and 73-71) and gSDSS being noisy for faint targets (e.g. 92-393 is marginal or noisy in indices involving g, but clearly variable in riz and H$\alpha$). The variability type is labeled as rotation ("Rot."), irregular ("Irr."),  dipper ("Dip."), burster ("Bur."), and uncertain ("Unc.").  }
    \label{tab:period-check}
\end{table*}

\section{Conclusions and outlook}
\label{conclu}

Using the first four months of data from the Javalambre North-PHASE Legacy Survey, we have explored and defined the criteria that should be applied in the identification of new young stars as well as other variable sources and performed a comprehensive study of one of the clusters, Tr~37. This initial part of the work allows us to demonstrate the power of the survey regarding the identification of YSO and other types of variables. The initial dataset, albeit limited, allows us to recover known variable stars in Tr~37, as well as to identify 50 new YSO candidates. 
We also show that the variability indices used give us a very good handle to detect different types of variability, with the Stetson indices being more sensitive to strong variables (e.g. those suffering accretion bursts or dips) and the SA04 indices allowing to detect milder but steadier variability, such as changes induced by cold or hot spots, as well as those related to accretion. 

Using the well-known Tr~37 cluster as a test-bench, we find that variability is a very good tool to identify new cluster candidate members, being moreover complementary to other techniques, including classical searches for disks and accretion signatures, and astrometry. Therefore, getting an integral and global picture of star formation within clusters can only be done using larger fields and a combination of techniques to avoid biases in the membership collection. 

Repeating the clustering analysis from PB23 including the variable stars strengthens the existing results, adding a significant number of younger, more variable stars in the surrounding clouds.
A clustering analysis including only variable stars also reveals substructure that had been missing before, such as the splitting the originally-monolithic main Tr~37 cluster into the central vs the globules population due to the higher degree of variability among objects in the surrounding globules. These differences, which had been washed out when using astrometry only, are a sign that variability can help to disentangle the star formation history to a higher degree than with Gaia data alone.
Being able to efficiently distinguish between types of variable stars using their astrometric properties is also a sign that we are finding a diverse non-young, variable population. 

Although the stellar properties are still beyond the reach of the survey, we are able to distinguish accreting objects as well as YSO with chromospheric activity. We also find significant differences between the variability indices for disked and diskless objects, particularly striking in the H$\alpha$ index. We observe that SA04 indices are more sensitive to small-scale variations, and there is also a clear difference between the performance of different bands, probably related to the diversity of variability causes (accretion, extinction, spots). Even if the strongest variability tends to be related to objects with massive protoplanetary disks as evidenced from their IR excesses, our combination of indices detects a similar fraction of variable YSO without disks in Tr~37. This evidences that our methodology can effectively detect variations from spots in young objects, which is further supported by the confirmation of rotational modulations and other types of variability for stars with magnitudes gSDSS$\sim$13-19 mag. 

North-PHASE finds other types of variables as well, including pulsating stars, eclipsing binaries, and long-period variables. We also see that most of the objects that have been suggested to be cluster members but were revealed as contaminants tend to be background variable giants, which interestingly also show different variability trends. Our analysis shows that the variability indices for field stars and objects that are the main source of contamination (e.g. those that appear in the same region of the HR diagram but are not cluster members) are significantly different from those of YSO. Investigating such differences in variability will help in the future to further distinguish YSO even when in the absence of further data or if astrometry is unreliable, as well as to assign YSO probabilities to new candidate sources based on variability alone.

From the results on Tr~37, we have now created a set of routines to identify young stars and examine their properties and can confirm that, so far, the North-PHASE survey exceeds its initial expectations, and will provide valuable data beyond the field of star formation.

\section*{Acknowledgements}

We thank the anonymous referee for comments that helped to clarify this paper.
We thank the Javalambre team for the time awarded to this project and their help with the planning and execution of the observations.
We also thank the OAJ Data Processing and Archiving Department (DPAD) for reducing and calibrating the OAJ data used in this work.
Based on observations made with the JAST80 telescope at the Observatorio Astrof\'{i}sico de Javalambre, in Teruel, owned, managed and operated by the Centro de Estudios de F\'{i}sica del Cosmos de Arag\'{o}n.

RK is funded by the STFC under grant number ST/W507404/1. LS was funded by the RAS Undergraduate Summer Bursary. VR acknowledges the support of the Italian National Institute of Astrophysics (INAF) through the INAF GTO Grant “ERIS \& SHARK GTO data exploitation” and the European Union’s Horizon 2020 research and innovation program and the European Research Council via the ERC Synergy Grant ``ECOGAL'' (project ID 855130). JCW and CM are Funded by the European Union (ERC, WANDA, 101039452). Views and opinions expressed are however those of the author(s) only and do not necessarily reflect those of the European Union or the European Research Council Executive Agency. Neither the European Union nor the granting authority can be held responsible for them. This work was also supported by the NKFIH excellence grant TKP2021-NKTA-64.

The Pan-STARRS1 Surveys (PS1) and the PS1 public science archive have been made possible through contributions by the Institute for Astronomy, the University of Hawaii, the Pan-STARRS Project Office, the Max-Planck Society and its participating institutes, the Max Planck Institute for Astronomy, Heidelberg and the Max Planck Institute for Extraterrestrial Physics, Garching, The Johns Hopkins University, Durham University, the University of Edinburgh, the Queen's University Belfast, the Harvard-Smithsonian Center for Astrophysics, the Las Cumbres Observatory Global Telescope Network Incorporated, the National Central University of Taiwan, the Space Telescope Science Institute, the National Aeronautics and Space Administration under Grant No. NNX08AR22G issued through the Planetary Science Division of the NASA Science Mission Directorate, the National Science Foundation Grant No. AST-1238877, the University of Maryland, Eotvos Lorand University (ELTE), the Los Alamos National Laboratory, and the Gordon and Betty Moore Foundation.
This work has made use of data from the European Space Agency (ESA) mission Gaia (\href{https://www.cosmos.esa.int/gaia}{https://www.cosmos.esa.int/gaia}), processed by the Gaia Data Processing and Analysis Consortium (DPAC, \href{https://www.cosmos.esa.int/web/gaia/dpac/consortium}{https://www.cosmos.esa.int/web/gaia/dpac/consortium}). Funding for the DPAC has been provided by national institutions, in particular the institutions participating in the Gaia Multilateral Agreement. 
This publication makes use of data products from the Wide-field Infrared Survey Explorer, which is a joint project of the University of California, Los Angeles, and the Jet Propulsion Laboratory/California Institute of Technology, funded by the National Aeronautics and Space Administration. This publication makes use of data products from the Wide-field Infrared Survey Explorer, which is a joint project of the University of California, Los Angeles, and the Jet Propulsion Laboratory/California Institute of Technology, funded by the National Aeronautics and Space Administration.
This paper includes data collected by the TESS mission. Funding for the TESS mission is provided by the NASA's Science Mission Directorate

This work is based in part on archival data obtained with the Spitzer Space Telescope, which was operated by the Jet Propulsion Laboratory, California Institute of Technology under a contract with NASA. Support for this work was provided by an award issued by JPL/Caltech.
This research has made use of the SIMBAD database,
operated at CDS, Strasbourg, France \citep{wenger00simbad}.
This research has made use of the VizieR catalogue access tool, CDS, Strasbourg, France. This research has made use of NASA’s Astrophysics Data System Bibliographic Services. 
This work has made use of TOPCAT and STILTS \citep{taylor05topcat} for the management of the data catalogues.
This work made use of Astropy:\footnote{\href{http://www.astropy.org}{http://www.astropy.org}} a community-developed core {\sc Python} package and an ecosystem of tools and resources for astronomy \citep{astropy_collaboration_astropy_2013,astropy_collaboration_astropy_2018}. 

\section*{Data Availability}

The North-PHASE project is a Legacy Survey at the Javalambre Observatory. As such, the data are released after publication in the Javalambre Observatory site\footnote{\href{https://archive.cefca.es/catalogues/north_phase-paper1 }{https://archive.cefca.es/catalogues/north\_phase-paper1 }}. On the long term, the project will contribute to the Legacy database at Javalambre and the Spanish Virtual Observatory, and all analysis tools, including customised {\sc Python} packages, will be released to the public via GitHub by the end of the project. In the meantime, the project team accepts reasonable requests of data and code and is open to collaboration.



\bibliographystyle{mnras}
\bibliography{biblio.bib} 



\appendix

 \section{Supplementary figures}
 \label{app-figs}

In this section we include some of the figures that complete or clarify parts of the work. Figure \ref{fig:compare-index-memb-nonmemb} shows the comparison between different types of variability indices, both for members and non-members, as discussed in Section \ref{basicvaria-sect}. Figure \ref{fig:index_vs_classes} presents the variability indices for objects in the variable groups as defined in Section \ref{variability-sect}. Figures \ref{fig:wise_vs_index} and \ref{fig:spitzer_vs_index_MIPS} show the infrared colours versus the variability indices for objects in the classes discussed in Section \ref{disk-sect}.

\begin{figure*}
    \centering
    \includegraphics[height=10cm]{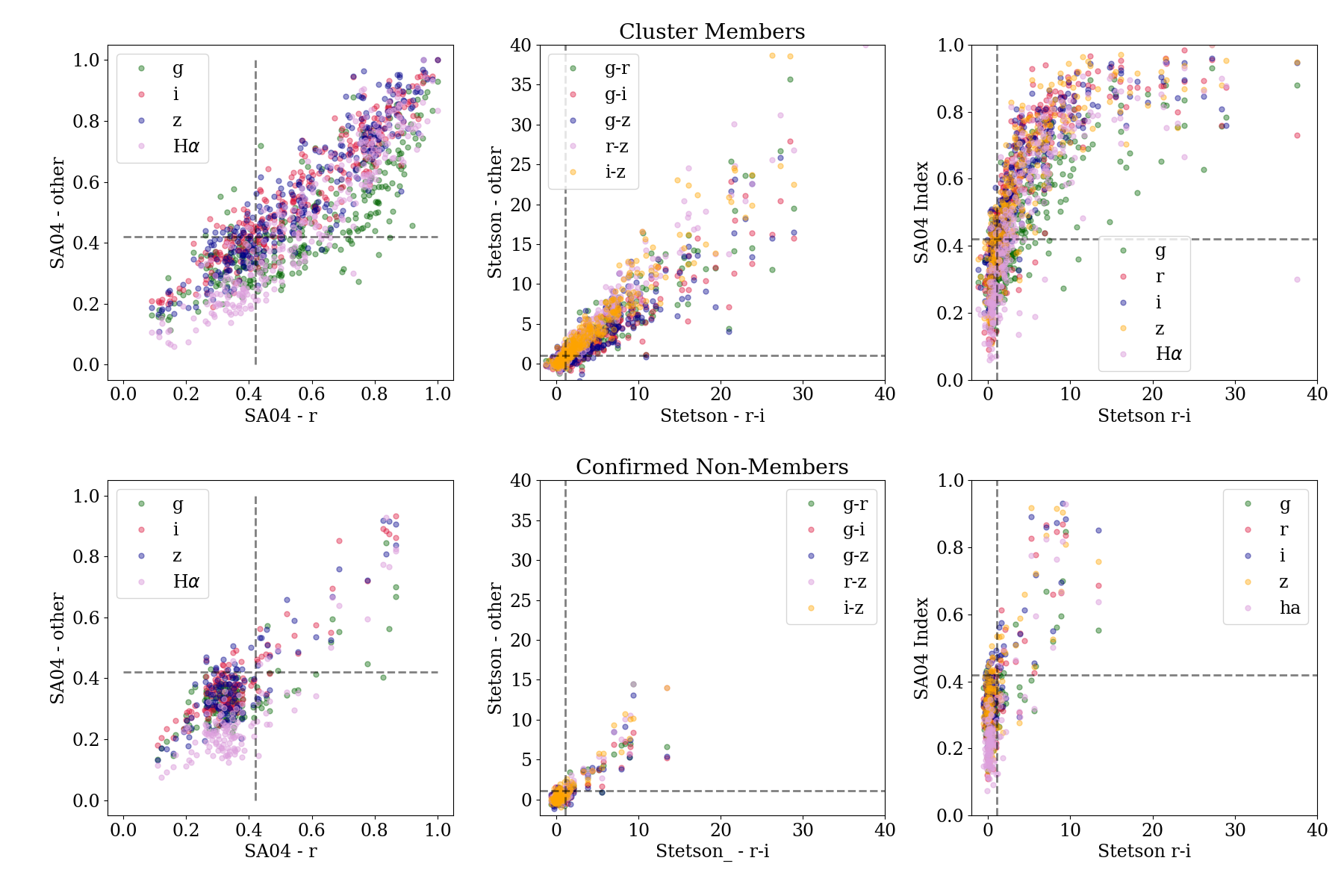}
    \caption{A comparison of various indices for confirmed members (to which we have also added the objects identified with Gaia; top panels) and the probably-non-member (bottom panel) from Table 1 in PB23. The dashed lines mark the limits beyond which we consider a star as variable (see Section \ref{variability-sect}). The variability indices are strongly correlated, and we find clear differences between objects known to be members vs those excluded from the membership class that share similar characteristics in terms of magnitude and colour. The number of variable objects is much higher among the highly reliable members, as expected, even though contaminants often include variable background giants. }
    \label{fig:compare-index-memb-nonmemb}
\end{figure*}

\begin{figure*}
    \centering
    \includegraphics[width=15cm]{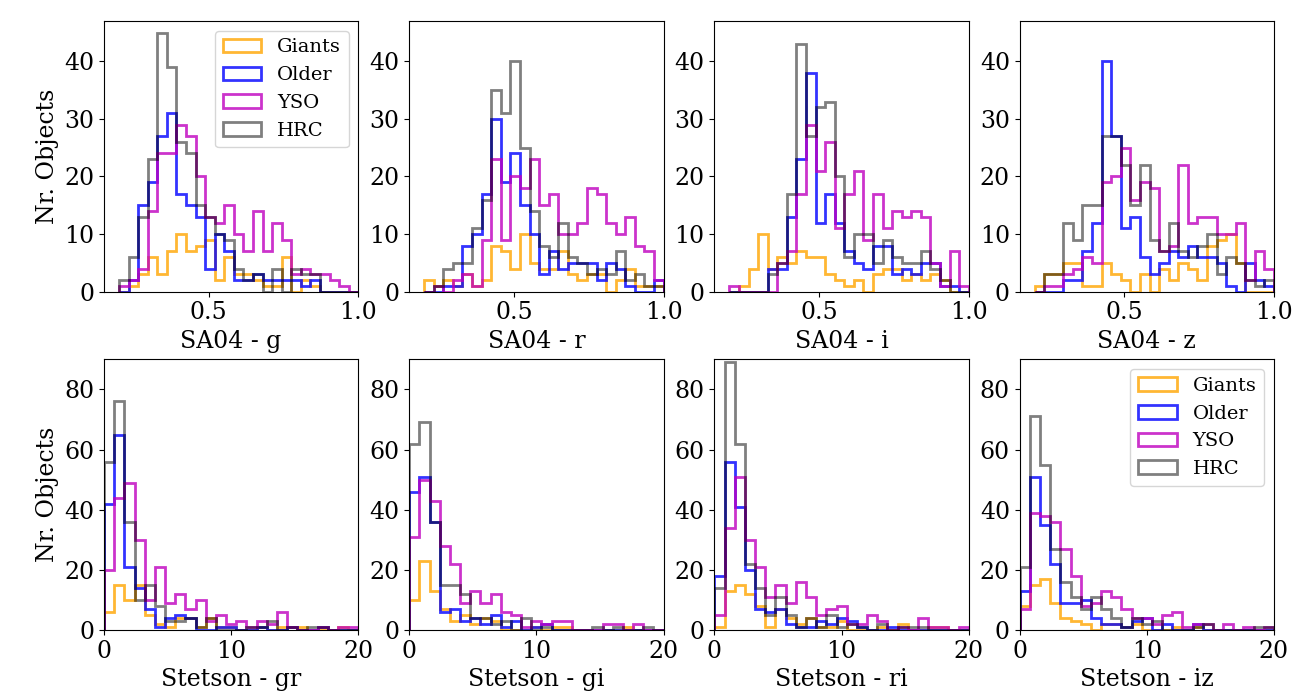}
    \caption{Individual variability indices for the different classes of variables as defined in Section \ref{variability-sect}. The groups refer to candidate giants, YSO, older variables, and HR contaminants. The differences between the various groups are evident.}
    \label{fig:index_vs_classes}
\end{figure*}

\begin{figure*}
    \centering
    \includegraphics[width=16cm]{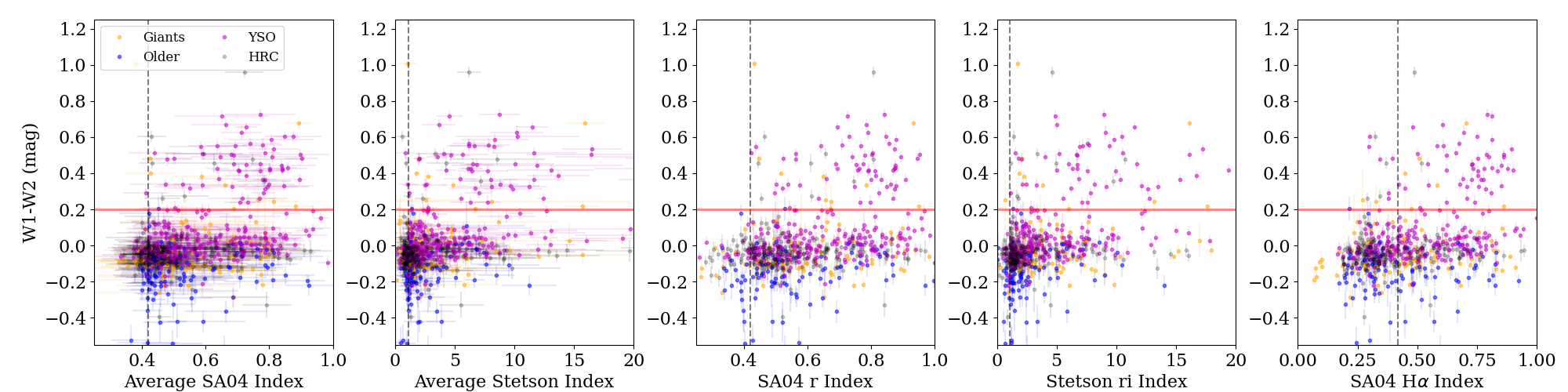}
        \includegraphics[width=16cm]{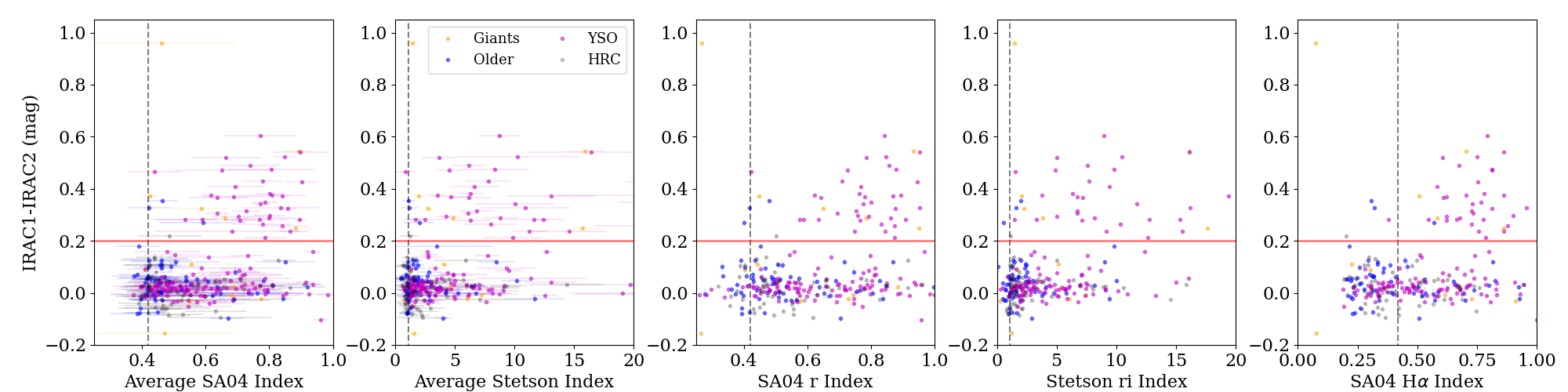}
    \caption{WISE W1-W2 (top) and Spitzer IRAC 1 - IRAC2 (bottom) colour for stars classified as variable. The type of variable star is the same than in Fig. \ref{fig:selection-new-variables}, and thus uses the parallax and position in the colour-magnitude diagram to find candidate YSO, Giants, older field stars, and objects that fall between the 0.5-20 Myr isochrones but do not appear to be at the cluster parallax ('HR contaminants' or HRC). The average SA04 and Stetson indices are shown (the errorbars are the standard deviation of the different indices), as well as the SA04 r and H$\alpha$ indices and the Stetson ri index. The vertical dashed line shows the limits considered for variability for each type of index. The red continuous horizontal line shows the lower limit for an object to be considered as having a disk (PB23). We note that, although many of the variable YSO candidate stars do have disks, not all the variable objects do. There is a trend for YSO with large variability indices to be disky, but a substantial number of objects with large variability have no near-IR excess.  }
    \label{fig:wise_vs_index}
\end{figure*}

\begin{figure*}
    \centering
    \includegraphics[width=16cm]{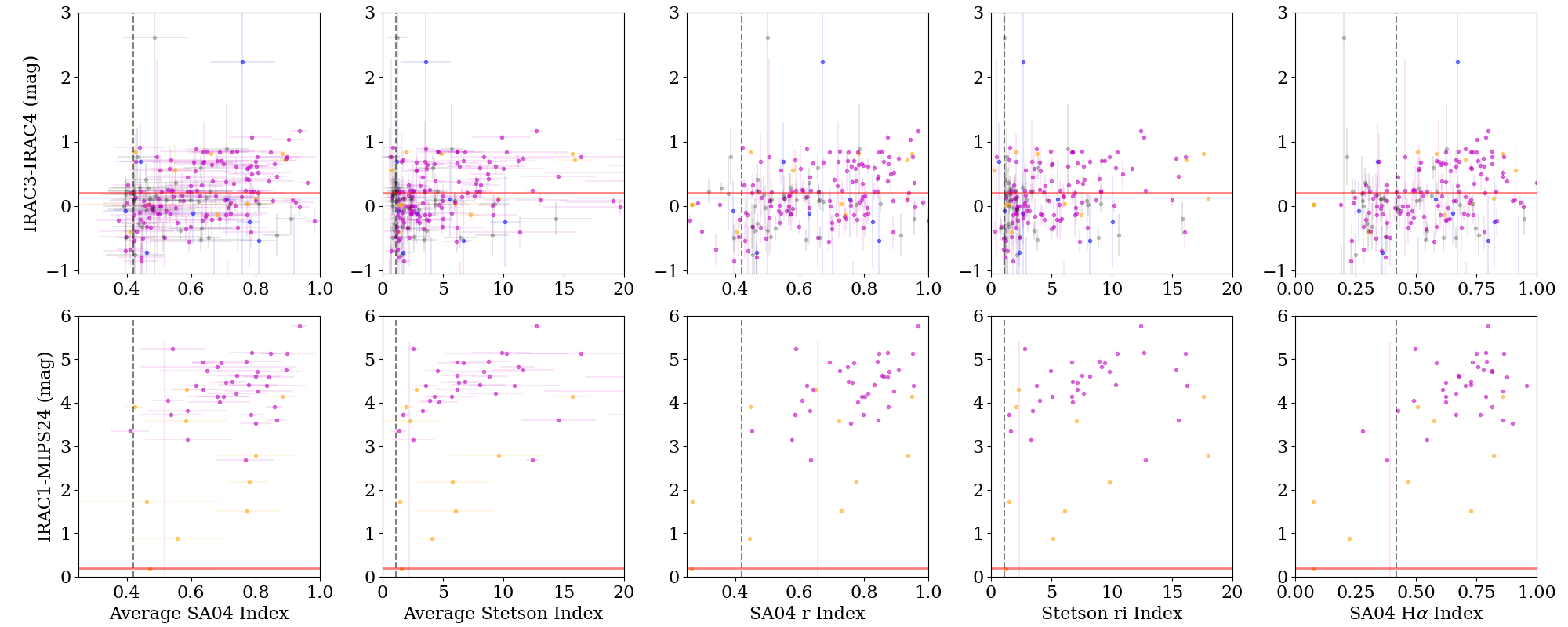}
    \caption{Same as Fig. \ref{fig:wise_vs_index}  but for Spitzer IRAC bands 3 and 4 (top) and MIPS 24$\mu$m (bottom). Note that the MIPS survey was only complete down to the photospheres of B-type stars at the distance of Tr~37, and hence all the objects detected in the bottom plots have infrared excesses, normally disks, and for that reason, they also tend to be highly variable.}
    \label{fig:spitzer_vs_index_MIPS}
\end{figure*}

\section{The Maximum Likelihood clustering procedure }
\label{likelihood-app}

The maximum likelihood algorithm is based on the technique deployed by \citet{Lindegren2000} and used by \citet{Roccatagliata2018,roccatagliata20} to study the structure of Taurus, and by PB23 to study the structure of the IC~1396 region, to which Tr~37 belongs. This technique explores the multidimensional astrometric space (considering parallax and the two proper motions) for a distribution of subclusters, assigning individual stars to subclusters and calculating the maximum likelihood for each set of subclusters. The algorithm uses a customised {\sc Python} routine and starts with an initial subcluster distribution, taken as a 'guess' from visually inspecting the distribution of objects. This guess is subsequently modified by the algorithm (typically, within a parameter space of $\pm$3$\sigma$) and the new likelihood is computed.  The final, most significant subcluster structure is determined by maximizing the likelihood function (see equations below).

In this multidimensional space, each j-th subcluster is defined by seven parameters: the mean parallax ($\varpi_{j}$), the mean proper motion in right ascension and declination ($\mu_{\alpha,j}$, $\mu_{\delta,j}$), their intrinsic dispersion for the stars in each subcluster ($\sigma_{\varpi,j}$, $\sigma_{\mu_\alpha,j}$, $\sigma_{\mu_\delta,j}$ ), and the fraction stars that belong to the subcluster ($fs_j$).
Conversely, for each i-th star, we consider its astrometric parameters parallax ($\varpi_i$) and proper motion in right ascension and declination ($\mu_{\alpha,i}$, $\mu_{\delta,i}$), with their corresponding uncertainties as well as correlated uncertainties, $\rho_i$ (note that the astrometric parameters are not independent). We then explore the 3-dimensional parameter space, where for the location of each star $i$ it is represented by the vector a$_i$=[$\varpi_i$, $\mu_{\alpha,i}$, $\mu_{\delta,i}$]. The positions of the stars are contrasted with the locations of the subclusters, so the astrometric positions for each j-th subcluster are given by the vector a$_{j}$=[$\varpi_{j}$, $\mu_{\alpha,j}$, $\mu_{\delta,j}$].

The probability of each i-th star belongs to the j-th subcluster is given by
\begin{equation}
P_{i,j}=fs_{j} \frac{L_{i,j}}{L_i}
\end{equation}
where $L_{i,j}$ is the likelihood function for the i-th star belonging to the j-th subcluster , $L_i$ is the total likelihood, and $fs_j$ is the fraction stars of belonging to the j-th subcluster. The individual likelihood $L_{i,j}$ is calculated as 
\begin{equation}
    L_{i,j}=(2\pi)^{2/3}|C_{i,j}|^{1/2}exp[-\frac{1}{2}(a_i-a_{j})^TC_{i,j}^{-1}(a_i-a_{j})],
    \label{eq_likelihood}
\end{equation}
and the sum of all individual likelihoods, multiplied by their corresponding fraction of stars, for the total of n subclusters, gives the total likelihood, i.e.
\begin{equation}
    L_i=\sum_{j=1}^n fs_jL_{i,j}.
\end{equation}
Equation \ref{eq_likelihood} can be understood as a multidimensional Gaussian distance, where $C_{i,j}$ is the covariance matrix, $|C_{i,j}|$ is its determinant, thus
\begin{equation}
C_{i,j}=
    \begin{bmatrix}
    C_{ij,11} & C_{ij,12}&C_{ij,13}\\
    C_{ij,21} & C_{ij,22}&C_{ij,23}\\
    C_{ij,31} & C_{ij,32}&C_{ij,33}\\
    \end{bmatrix}
\end{equation}
where the different terms of the mattrix correspond to the various uncertainties, including correlated terms, i.e.,
\begin{multline}
[C_{ij,11}] = \sigma^2_{\varpi,i} + \sigma^2_{\varpi,j}\\
[C_{ij,22}] = \sigma^2_{\mu_\alpha,i} + \sigma^2_{\mu_\alpha,j}\\
[C_{ij,33}] = \sigma^2_{\mu_\delta,i} + \sigma^2_{\mu_\delta,j}\\
[C_{ij,12}] = [C_{ij,21}]= \sigma_{\varpi,i}\times \sigma_{\mu_\alpha,i}\times \rho_i(\varpi,\mu_\alpha)\\
[C_{ij,13}] = [C_{ij,31}]= \sigma_{\varpi,i}\times \sigma_{\mu_\delta,i}\times \rho_i(\varpi,\mu_\delta)\\
[C_{ij,23}] = [C_{ij,32}]= \sigma_{\mu_\alpha,i}\times \sigma_{\mu_\delta,i}\times \rho_i(\mu_\alpha,\mu_\delta).\\
\end{multline} 
In addition, $a_i$ is the multidimensional vector for the i-th star, $a_i$=[$\varpi_{i}$ $\mu_{\alpha,i}$ $\mu_{\delta,i}$], and $a_{j}$ is the equivalent vector for the subcluster, $a_{j}$=[$\varpi_{j}$ $\mu_{\alpha,j}$ $\mu_{\delta,j}$], and $^T$ indicates transposition. 
Therefore,
\begin{equation}
a_i-a_{j}=
    \begin{bmatrix}
    \varpi_i - \varpi_{j}\\
    \mu_{\alpha,i} - \mu_{\alpha,j}\\
    \mu_{\delta,i} - \mu_{\delta,j}\\
    \end{bmatrix}
\end{equation}
The total likelihood is then computed as the sum of all likelihoods for all the stars. In our case and for the total population, the maximum likelihood is 823.4.

\section{A note on accretion rates from H$\alpha$ fluxes}
\label{accretion-app}

Measuring accretion rates is in general complicated, since a very good knowledge of the source is required \citep[e.g.][]{sicilia10} and, moreover, numerous calibrations exist regarding the conversion between excess luminosity (in photometry or lines) and accretion luminosity \citep[e.g.][]{alcala_x-shooter_2014,Fang2009}. In general, the more indirect the method is, the higher the risk of miscalibrations \citep{mendigutia15}. Ahead of the North-PHASE complete survey, we explored the existing correspondence between accretion rates from U band and H$\alpha$, using the data for Tr~37. 
Figure \ref{fig:havsuband} shows a comparison between the values from \citet{sicilia10}, obtained via U-band excess, and those from \citet{barentsen11}, derived from H$\alpha$ photometry, which reveals a systematic offset, the values from H$\alpha$ being systematically higher by a factor of few, compared with those derived from U-band .

To check if the difference lay in the way the H$\alpha$ EW is estimated from photometry vs the spectral value, we compared the H$\alpha$ EW from \citet{sicilia06hecto,sicilia13,nakano12}, measured from spectra, and those from photometry \citep{barentsen11}. This reveals no significant differences, albeit a large level of variability for individual objects, which had been also documented when comparing the spectra themselves \citep{sicilia06hecto,sicilia13}. The results are shown in Fig. \ref{fig:hacomp}. Nevertheless, the depth of H$\alpha$-based and U-excess-based measurements of the accretion rate is not comparable, so variable accretion together with detection limits may induce at least part of this behaviour. 

A similar offset is so far found when estimating H$\alpha$ EW from the North-PHASE photometry (see Fig. \ref{fig:haEW}). In this case, a strong limitation lies in the lack of information on the individual spectral types and extinctions, which will be a particularly large problem for red stars (both due to extinction or to effective temperature). Since North-PHASE will offer constraints on the stellar properties once variability is well-determined, this issue will be addressed and further explored in the future.

\begin{figure}
    \centering
    \includegraphics[width=7cm]{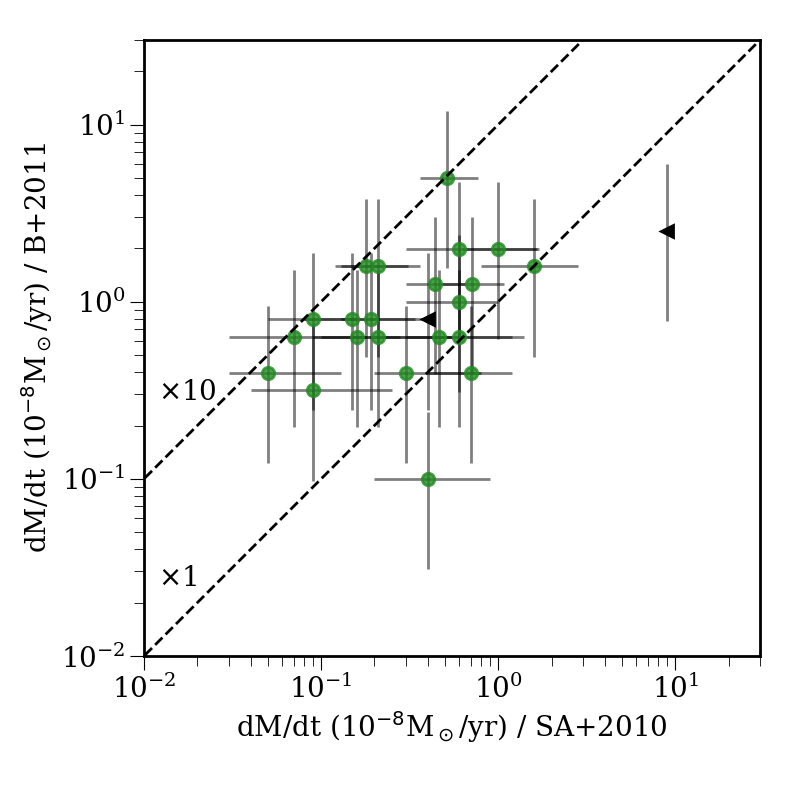}
    \caption{A comparison of the accretion rates as derived from u-band photometry \citep[x-axis,][]{sicilia10} and H$\alpha$ photometry \citep[y axis,][]{barentsen11}. The generalised offset of accretion rates calculated from photometry and from U-band excesses reveals that we need to be cautious with the calibration.  }
    \label{fig:havsuband}
\end{figure}

\begin{figure}
    \centering
    \includegraphics[width=7cm]{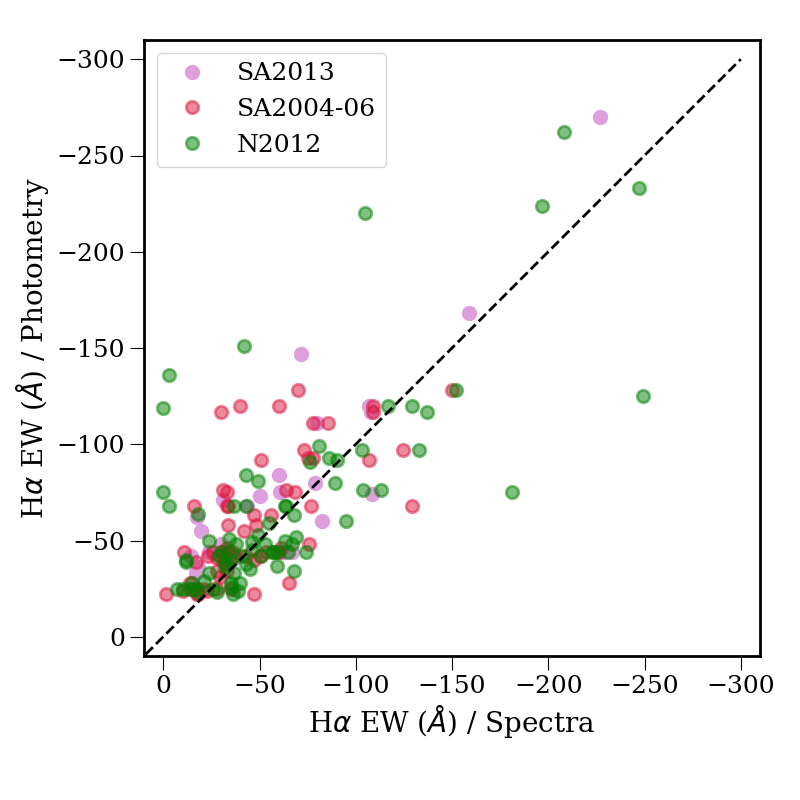}
    \caption{A comparison of H$\alpha$ EW measured from spectra \citep[x-axis,][]{sicilia06hecto,sicilia13,nakano12} vs IPHAS photometry \citep[y-axis][]{barentsen11}. The dashed line shows the identity result. Although there are significant differences between measurements of the same object taken on different dates, even if the same method is used, and there is a slight tendency for weak H$\alpha$ to be missed by photometry, the agreement is reasonable.  }
    \label{fig:hacomp}
\end{figure}

\begin{figure}
    \centering
    \includegraphics[width=8cm]{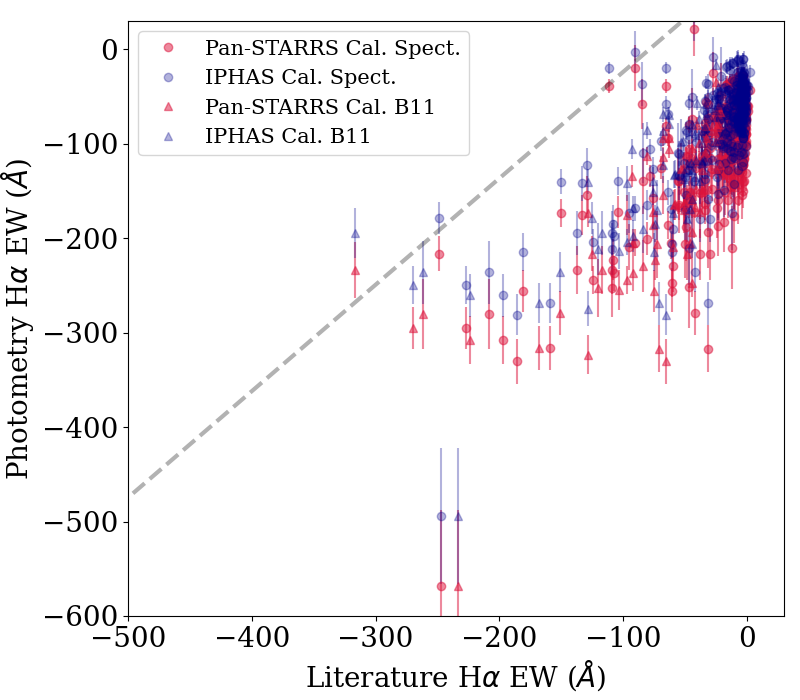}
    \caption{A comparison of the first-check H$\alpha$ EW from North-PHASE. 
    The x axis contains measurements from the literature (see text, circles correspond to spectral measurements, while triangles denote photometric estimates). In red, we show the results using Pan-STARRS data for calibration of the rSDSS filter, while in blue we show the same results but using rSDSS from IPHAS to calibrate our rSDSS band. The H$\alpha$ data are always calibrated with IPHAS. Although this 'rough estimate' is clearly affected by systematics, we find a clear correlation which demonstrates that the North-PHASE J0660 photometry can be used to estimate the strenght of the H$\alpha$ line.}
    \label{fig:haEW}
\end{figure}

Regarding the accretion rates, lower rates seem more consistent in general, especially because very high rates start being in conflict with disk masses and stellar ages \citep{spezzi12,sicilia-aguilar_`rosetta_2016}. On the other hand, the H$\alpha$ measurements from spectra and photometry appear to be very robust and, notwithstanding the source variability, indistinguishable. The reasons for the difference in accretion rates could be related to issues with the calibration, since line fluxes are always more uncertain than UV excesses to get accretion, and even more if they come from photometry. 
But, since the probed range of accretion rates is not too different from the typical factor-of-few variability observed in individual objects \citep[and from the typical uncertainties; ][]{sicilia10} there is also the possibility of selection effects, since H$\alpha$ photometric measurements pick more extreme cases/objects thus favouring the detection of objects when they are at a higher accretion state.
Although at this stage it is difficult to find a cause for these differences, the variability aspect of North-PHASE will offer a great opportunity to track the changes in H$\alpha$ photometry and improve the calibration in the future.

\end{document}